\documentclass[aps,prl,amsfonts,floatfix,superscriptaddress,longbibliography,twocolumn,footinbib]{revtex4-2}

\usepackage[utf8]{inputenc}
\usepackage{amsmath}    
\usepackage{amssymb}
\usepackage{mathbbol}  
\usepackage{verbatim}   
\usepackage{subfigure} 
\usepackage{scalerel}
\usepackage{soul}
\usepackage{siunitx}
\usepackage{physics}
\usepackage{enumerate} 
\usepackage{empheq}

\usepackage{graphicx}% Include figure files
\usepackage{dcolumn}% Align table columns on decimal point
\usepackage{bm}% bold math
\usepackage[colorlinks = True, linkcolor = red, citecolor = red]{hyperref}
\usepackage{graphicx}
\usepackage[normalem]{ulem}
\usepackage{xcolor}
\usepackage{enumitem}
\usepackage{orcidlink}

\usepackage{tikz}
\usepackage{pgfplots}
\usetikzlibrary{shapes, arrows, positioning, calc}
\pgfplotsset{compat=1.18}

\definecolor{panelblue}{HTML}{002D62}
\definecolor{panelgreen}{HTML}{004B23}
\definecolor{panelpurple}{HTML}{4A0E4E}
\definecolor{panelorange}{HTML}{D9541E}

\definecolor{curveblue}{HTML}{1A73E8}
\definecolor{curvered}{HTML}{D93025}
\definecolor{curvegreen}{HTML}{1E8E3E}

\tikzset{
    axis/.style={thick, ->, >=stealth, black},
    label/.style={font=\fontfamily{phv}\selectfont\small},
    sublabel/.style={font=\fontfamily{phv}\selectfont\footnotesize\itshape}
}
\begin{document}

\title{Observable- and state-selective prethermalization and bounds on prethermal lifetimes}

\author{C. L. Sriram\,\orcidlink{0009-0001-3706-6498}} 
%\email{c.l.sriram@uconn.edu}
\affiliation{Department of Physics, University of Connecticut, Storrs, Connecticut, USA}

\author{Soumya Kanti Pal\,\orcidlink{0009-0008-7226-356X}} 
%\email{soumya.pal@tifr.res.in} 
\affiliation{Department of Theoretical Physics, Tata Institute of Fundamental Research, Homi Bhabha Road, Mumbai 400005, India}

\author{Lea F. Santos\,\orcidlink{0000-0001-9400-2709}} 
%\email{lea.santos@uconn.edu} 
\affiliation{Department of Physics, University of Connecticut, Storrs, Connecticut, USA}

\begin{abstract}
Prethermalization describes long-lived intermediate regimes that precede equilibrium and can dominate experimentally accessible dynamics. Here, we show that a separation of spectral energy scales, despite giving rise to a hierarchy of dynamical timescales, does not by itself guarantee the appearance of a prethermal plateau. Under the same Hamiltonian, some observables may exhibit prethermal behavior,  while others relax directly toward equilibrium. We uncover the mechanism governing this selectivity, showing that the emergence of a prethermal plateau depends not only on the Hamiltonian, but also on the observable and the initial state. Our results apply broadly to systems close to a fully permutation-symmetric limit and are illustrated with a long-range interacting spin model. We further prove that the Loschmidt echo provides a lower bound on the prethermal lifetime of any bounded observable whenever the prethermal and exact dynamics are unitary. 
\end{abstract}

\maketitle

Prethermalization refers to a long-lived intermediate dynamical regime that precedes the eventual equilibration of a quantum system. The lifetime of prethermal states can be so long, that much of the experimentally accessible dynamics may occur within the prethermal regime rather than near the equilibrium state.  Prethermalization has been found in a broad range of settings~\cite{Barnett2011,Kollar2011,Babadi2015,Alba2017,Abanin2017,Mori2018,Reimann2019_01,Yin2023,Gallone2026}, including systems with weakly integrability breaking perturbations~\cite{Bertini2015,Else2017,Mallayya2019,Mallayya2021}, periodically and aperiodically driven systems~\cite{Kuwahara2016,Ho2018,Dumitrescu2018,Machado2019,Machado2020,Mori2021,Santos2021,Bhakuni2021,Das2023,Dea2024,Tiwari2025}, systems with long-range interactions~\cite{Kastner2010,Schutz2014,Schutz2016,Mori2019,Defenu2024_1,ArrufatVicente2025,Manju2026}, and far-from-equilibrium quantum field theories~\cite{Halimeh2020,Hayata2024}. Experimental observations span several of these settings, including weakly perturbed integrable systems~\cite{Tang2018}, long-range interacting spin systems~\cite{Neyenhuise2017}, and driven interacting systems~\cite{Wei2019,Peng2021,Rubio2020}. More recently, programmable quantum processors have enabled the observation of remarkably long-lived and controllable prethermal dynamics~\cite{Bao2026,Liu2026}.

The mechanisms responsible for prethermalization depend on the physical setting. In driven systems, rapid or structured driving can suppress energy absorption and postpone heating, enabling prethermal regimes governed approximately by an effective Hamiltonian. In isolated systems described by time-independent Hamiltonians, prethermalization can instead arise from effects that are often interconnected and include weakly broken integrability or symmetries and hierarchies of energy scales that generate widely separated dynamical timescales. Regardless of their microscopic origin, a central question concerns the lifetime of the resulting prethermal regime, with considerable efforts devoted to identifying and controlling prethermal timescales.

However, before addressing how long a prethermal plateau survives, there is an even more fundamental question: Does a Hamiltonian that supports a prethermal regime necessarily produce a prethermal plateau in the dynamics of an arbitrary observable? We show that the existence of well-separated dynamical timescales alone does not guarantee that they will be manifested in every quantity used to probe the system. The dynamics depends not only on the spectrum of the Hamiltonian, but also on the observable and the initial state. Our studies demonstrate that prethermalization is selective. Under the same Hamiltonian, one observable may display a long-lived prethermal plateau while another relaxes directly toward equilibrium; and even for the same observable, a prethermal plateau may emerge for one initial state but be absent for another. We identify the mechanism underlying this observable- and state-selective prethermalization.

We demonstrate this selectivity for a broad class of Hamiltonians in which weak breaking of full permutation symmetry results in a many-body spectrum organized into nearly degenerate energy bands. The distinct interband and intraband energy scales generate well-separated dynamical timescales. We derive a symmetry-based selection rule that excludes prethermal plateaus for observables preserving full permutation symmetry, while for observables not subject to this constraint, their emergence depends on the initial state's overlap with the relevant intraband states. As a representative realization, we study an experimentally relevant spin-1/2 model with strong long-range interactions.

Having established when a prethermal plateau emerges, we turn to the question of how long it persists. We prove that the Loschmidt echo provides a lower bound on the prethermal lifetime of any bounded observable. The bound relies only on the unitary nature of the exact and prethermal dynamics and therefore applies to any system admitting a unitary prethermal description, including periodically driven systems.

%%%%%%%%%%%%%%%%%%%%%%%%%%%%%%%%%%%%%
%%%%%%%%%%%%%%%%%%%%%%%%%%%%%%%%%%%%%
%%%%%%%%%%%%%%%%%%%%%%%%%%%%%%%%%%%%%
{\em Model and energy bands.--} We consider the following one-dimensional spin-$1/2$ transverse field Ising model with power-law interactions, as realized in experiments with ion traps~\cite{Jurcevic2014,Richerme2014,Neyenhuise2017},
\begin{align}
\hat{H}^{(\alpha)} = \sum_{i<j} \frac{J}{\mathcal{N}_\alpha} \frac{\hat{\sigma}_i^x \hat{\sigma}_j^x}{|i-j|^\alpha}
+ h \sum_{i=1}^{L} \hat{\sigma}_i^z,
\label{eq:XXtZH}
\end{align}
where $\hat{\sigma}_i^{x,y,z}$ are Pauli matrices at site $i$, $L$ is the system size, $J=1$ sets the interaction scale, $h=1$ is the transverse magnetic field strength, and $\mathcal{N}_\alpha \sim L^{1-\alpha}$ is the Kac's factor. The Hamiltonian is invariant under parity and spin inversion. 

\begin{figure}[h]
    \begin{center}
        \includegraphics[width = 1.0\columnwidth]{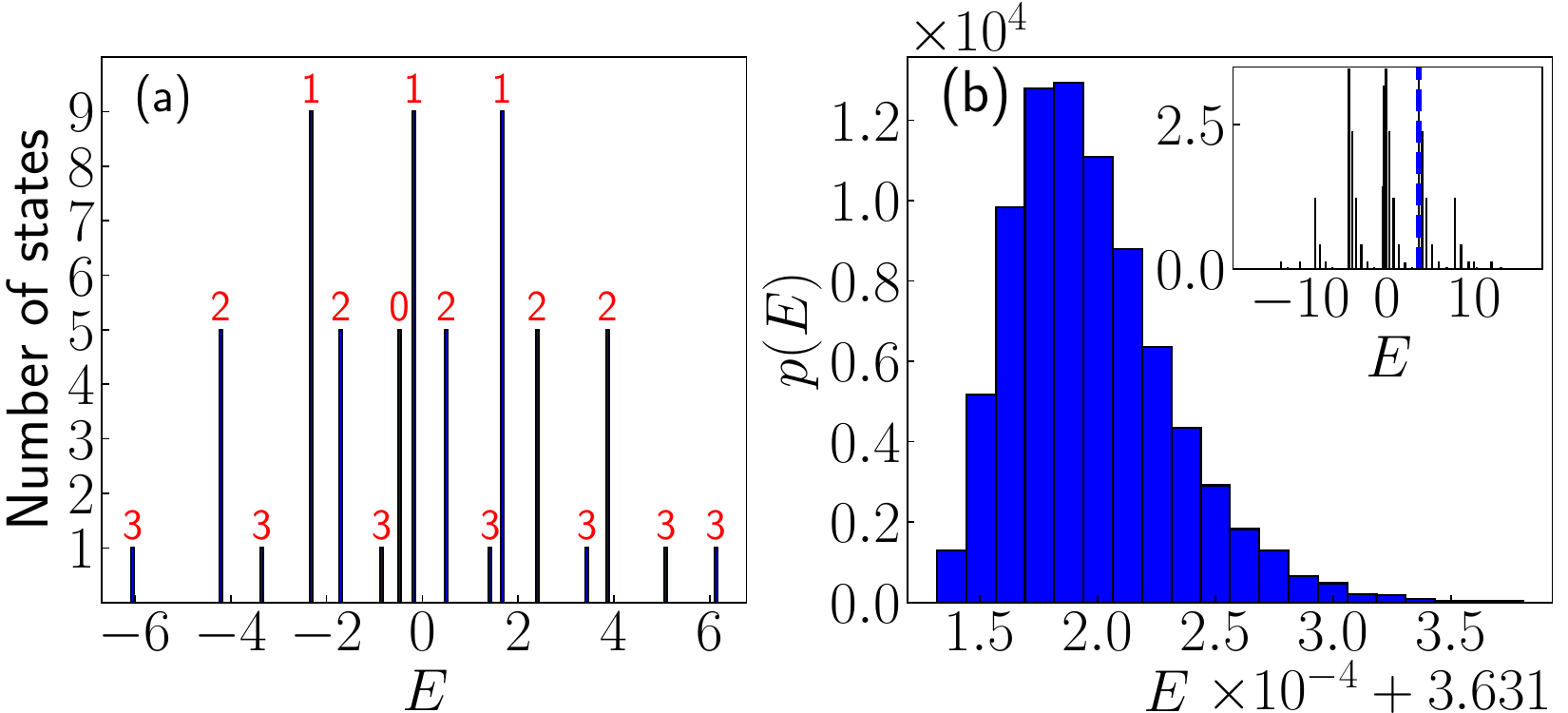}
        \caption{Density of states for (a) $\hat{H}^{(0)}$ with $L=6$ and $s$ for each band indicated in red, and (b) a single energy band of $\hat{H}^{(\alpha \neq 0)}$ with $L=16$ and $\alpha=10^{-4}$.  The inset shows that the full spectrum of $\hat{H}^{(\alpha \neq 0)}$ retains the band structure. 
        }
        \label{fig:FIG1}
    \end{center}
\end{figure}

In the all-to-all limit, $\alpha=0$, Eq.~(\ref{eq:XXtZH}) reduces to the Lipkin-Meshkov-Glick (LMG) Hamiltonian, $\hat{H}^{(0)}$, an experimentally realizable collective-spin system~\cite{Li2023,Luo2025}. Its full permutation symmetry organizes the many-body spectrum into highly degenerate energy bands labeled by the total spin $s$, as shown by the density of states (DOS) in Fig.~\ref{fig:FIG1}(a). For any $\alpha>0$, permutation symmetry is broken and these degeneracies are lifted [Fig.~\ref{fig:FIG1}(b)]. However, in the strong-long-range regime, $0<\alpha<1$, as considered in this work, the band structure inherited from the fully connected limit persists [inset of Fig.~\ref{fig:FIG1}(b)]. Despite this banded structure, the spectral statistics within individual energy bands exhibit signatures of quantum chaos~\cite{Russomanno2021,Pal2025}, indicating that the system must thermalize at long times~\cite{Pal2025,Sriram2026}.

%%%%%%%%%%%%%%%%% FIGURE %%%%%%%%%%%
%%%%%%%%%%%%%%%%%%%%%%%%%%%%%%%%%%%%%

In the regime $0 < \alpha \ll 1$, the Hamiltonian can be viewed as a perturbation of the fully connected limit,
\begin{align}
\hat{H}^{(\alpha)} = \hat{H}^{(0)} + \epsilon \hat{V}_\alpha ,
\label{Eq:PertH_H0_V}
\end{align}
where $\hat{V}_\alpha$ introduces nonlocal corrections to the fully connected model and $\epsilon$ controls the perturbation strength. We denote the eigenvalues and eigenstates of the perturbed and fully connected Hamiltonians, respectively, by $\hat{H}^{(\alpha\neq 0)} |n \rangle = E_n |n \rangle $ and $ \hat{H}^{(0)} |n^{(0)} \rangle = E_n^{(0)} |n^{(0)} \rangle$.

%%%%%%%%%%%%%%%%%%%%%%%%%%%%%%%%%%%%%
%%%%%%%%%%%%%%%%%%%%%%%%%%%%%%%%%%%%%
%%%%%%%%%%%%%%%%%%%%%%%%%%%%%%%%%%%%%
{\em Two-stage equilibration process.--} We study the dynamics following a quantum quench. The system is initially prepared in an eigenstate of $\hat{H}_{\text{ini}} = \sum_{i=1}^L \hat{\sigma}_i^z$, so that the initial state $|\Psi_0\rangle$ corresponds to a product state in the $z$-basis.

%%%%%%%%%%%%%%%%% FIGURE %%%%%%%%%%%
\begin{figure*}
    \begin{center}
        \includegraphics[width = 2.0\columnwidth]{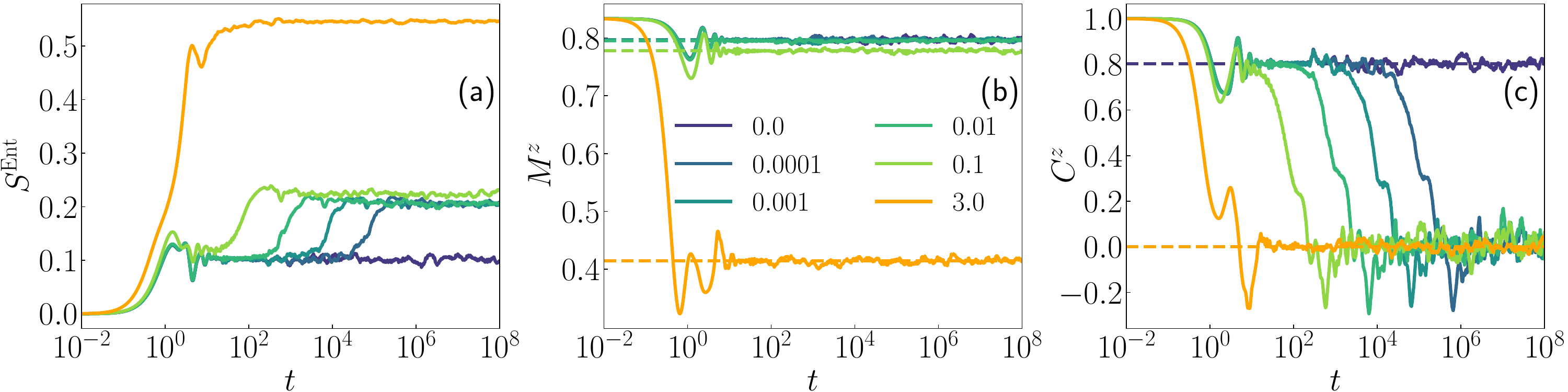}
        \caption{Evolution of the (a) entanglement entropy, (b)  magnetization along the $z$-direction $M^z$ and (c) first moment of the excitation density $C^z$ for an initial product state and different interaction-range exponents $\alpha$, as indicated in panel (b). The magnetization does not exhibit a prethermal plateau. 
      }
        \label{fig:FIG02}
    \end{center}
\end{figure*}
%%%%%%%%%%%%%%%%%%%%%%%%%%%%%%%%%%%%%

Figure~\ref{fig:FIG02}(a) shows the evolution of the von Neumann  entanglement entropy,
\begin{align}
    S^\mathrm{Ent} (t) &= \frac{-2}{L \log 2} \mathrm{Tr}_{L/2} \biggl[ \hat{\rho}_{L/2} (t) \log \hat{\rho}_{L/2} (t) \biggr],
    \label{Eq:Sent}
\end{align}
where $\hat{\rho}_{L/2}(t)$ is the reduced density matrix obtained by tracing out half of the system. We consider different values of the interaction-range exponent $\alpha$. Unlike the short-range case ($\alpha=3$) or the fully connected limit ($\alpha=0$), the entropy in the super-long-range regime exhibits a prethermal plateau whose lifetime increases as $\alpha$ decreases. This two-stage equilibration reflects the well-separated dynamical timescales generated by the band structure of the many-body spectrum.
%This two-stage equilibration originates from the band structure of the many-body spectrum. When expanded in the eigenbasis of $\hat{H}^{(\alpha)}$, the initial state has support across multiple energy bands, giving rise to two well-separated dynamical processes: rapid interband dephasing, which drives $S^\mathrm{Ent} (t)$ to the prethermal plateau, followed by much slower intraband dynamics that ultimately leads to the final equilibrium.

To investigate whether the two-stage process is also manifested in experimental local observables,
Figs.~\ref{fig:FIG02}(b) and \ref{fig:FIG02}(c) show, respectively, the dynamics of the total magnetization along the $z$-axis,
\begin{align}
    M^z(t) &= \frac{1}{L} \sum_{i=1}^{L}
    \langle \Psi(t)| \hat{\sigma}_i^z | \Psi(t) \rangle.
    \label{Eq:Mz}
\end{align} 
and the first moment of the excitation density introduced in the experimental study of prethermalization in long-range interacting spin chains~\cite{Neyenhuise2017},
\begin{align}
    C^z(t)
    &= \sum_{i=1}^{L}
    \langle\Psi(t)|x_i\hat{n}_i|\Psi(t)\rangle,
    \label{Eq:Cz}
\end{align}
where
$x_i=(2i-L-1)/(L-1)$
is the normalized position of site $i$ and
$\hat{n}_i=(\hat{\sigma}_i^z+\hat{\mathbb I})/2 $
is the local number operator.

In contrast to $C^z(t)$ [Fig.~\ref{fig:FIG02}(c)], which, like the entanglement entropy, exhibits a prethermal plateau, the magnetization $M^z$ [Fig.~\ref{fig:FIG02}(b)] relaxes directly toward equilibrium. We show next that this qualitative difference can be understood within a perturbative framework.

%%%%%%%%%%%%%%%%%%%%%%%%%%%%%%%%%
%%%%%%%%%%%%%%%%%%%%%%%%%%%%%%%%%
%%%%%%%%%%%%%%%%%%%%%%%%%%%%%%%%%
{\em Conditions for prethermal plateaus.--} To leading order in the perturbation strength $\epsilon$, the time evolution of the expectation value of an observable $\hat O$ is given by
\begin{align}
    \langle \hat{O} \rangle (t) &= \!\! \!\! \sum_{\substack{m \neq n \\ E_m^{(0)} \neq E_n^{(0)}}} \!\!\!\! \! \!{c_m^*}^{(0)} c_n^{(0)} \langle m^{(0)}|\hat{O}|n^{(0)} \rangle e^{i t[(E_m^{(0)} - E_n^{(0)}) 
    + \mathcal{O}(\epsilon)]} \nonumber \\
    & + \!\!\!\! \sum_{\substack{m \neq n \\ E_m^{(0)} = E_n^{(0)}}} \!\! \!\! \! \! {c_m^*}^{(0)} c_n^{(0)} \langle m^{(0)}|\hat{O}|n^{(0)} \rangle e^{i\epsilon t(E_m^{(1)} - E_n^{(1)})} \nonumber \\
    & + \sum_n |c_n^{(0)}|^2\langle n^{(0)}|\hat{O}|n^{(0)} \rangle  ,
    \label{eq:Obs_pre}
\end{align}
where $c_n^{(0)}=\langle n^{(0)}|\Psi_0\rangle$ and $E^{(1)}_n$ is the first-order correction in perturbation theory to $E_n^{(0)}$. The last term in Eq.~(\ref{eq:Obs_pre}) is the zeroth-order approximation to the long-time, or diagonal-ensemble, expectation value, 
$\langle \hat{O} \rangle^{\rm DE}
=
\sum_n |c_n|^2\langle n|\hat{O}|n\rangle$. 
The first two terms in Eq.~(\ref{eq:Obs_pre}) evolve on well-separated timescales. The first sum couples states belonging to different energy bands, whose large energy differences generate rapidly oscillating phases and therefore fast dephasing. The second sum couples states that are degenerate in $\hat H^{(0)}$, so their dynamics is governed by the much smaller intraband splittings and occurs on a correspondingly longer timescale. 

In the intermediate-time regime, after the interband contributions have dephased but before appreciable intraband dephasing occurs, Eq.~(\ref{eq:Obs_pre}) gives the prethermal plateau,
\begin{align}
    \langle \hat{O} \rangle_\mathrm{pre} &= \sum_{\substack{m \neq n \\ E_m^{(0)} = E_n^{(0)}}} \!\! \!\! \! \! {c_m^*}^{(0)} c_n^{(0)} \langle m^{(0)}|\hat{O}|n^{(0)} \rangle \nonumber \\
    & + \sum_n |c_n^{(0)}|^2\langle n^{(0)}|\hat{O}|n^{(0)} \rangle  .    \label{eq:Obs_pre2}
\end{align}
Its distinction from the long-time saturation value
$\langle \hat{O} \rangle^{\rm DE}$ requires that two conditions are simultaneously satisfied: (i) the observable has non-vanishing off-diagonal matrix elements $\langle m^{(0)}|\hat O|n^{(0)}\rangle$ between states that remain degenerate under $\hat H^{(0)}$, and (ii) the initial state has finite overlap with those states. 
If either condition fails, the prethermal plateau is not visible.

As proven in the Supplemental Material (SM)~\cite{noteSUPPL}, observables invariant under the full permutation group cannot distinguish states $|n^{(0)}\rangle$ that differ only in their multiplicity within a given energy band. Their off-diagonal matrix elements between such states therefore vanish, precluding a prethermal plateau. This class includes collective observables, such as the total magnetization along any direction.

%%%%%%%%%%%%%%%%% FIGURE %%%%%%%%%%%
\begin{figure}[h!]
    \begin{center}
        \includegraphics[width = 0.9\columnwidth]{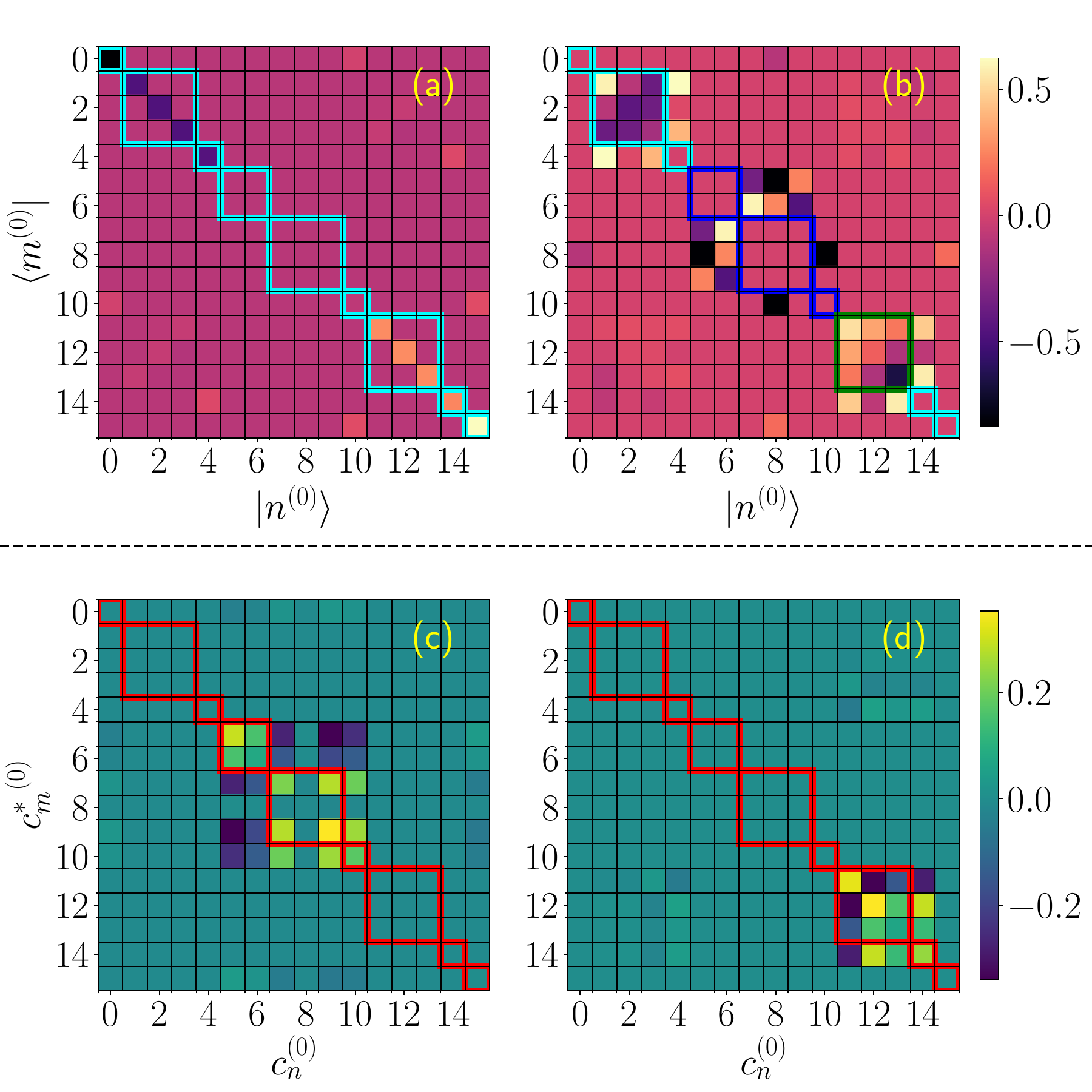}
        \caption{Matrix elements of (a) the permutation-invariant magnetization $\hat{M}^z$ and (b) the first moment of the excitation density $\hat{C}^z$ in the energy eigenbasis of $\hat{H}^{(0)}$. 
        Overlap matrices ${c_m^*}^{(0)}c_n^{(0)}$ for the initial state (c) $|\uparrow \downarrow \downarrow \uparrow \rangle$ and (d) $|\downarrow \downarrow \uparrow \downarrow \rangle$. 
Cyan, blue, green, and red boxes delimit degenerate energy bands.
        The blue and green boxes in (b) identify the bands relevant to the initial states in (c) and (d), respectively.
        }
        \label{fig:FIG03permut}
    \end{center}
\end{figure}
%%%%%%%%%%%%%%%%%%%%%%%%%%%%%%%%%

Figure~\ref{fig:FIG03permut}(a) illustrates this result for the magnetization $\hat M^z$, showing its matrix elements in the eigenbasis of $\hat H^{(0)}$. The cyan boxes delimit the degenerate energy bands, and the absence of off-diagonal matrix elements within each box confirms that $\hat M^z$ cannot couple states belonging to the same band. The first term in Eq.~(\ref{eq:Obs_pre}) therefore vanishes, explaining the absence of a prethermal plateau in $M^z(t)$ [Fig.~\ref{fig:FIG02}(b)].

By contrast, $\hat C^z$ is not permutation invariant and possesses finite off-diagonal matrix elements within several degenerate energy bands (blue, green, and cyan boxes), as seen in Fig.~\ref{fig:FIG03permut}(b). It therefore satisfies the first condition for a prethermal plateau. Whether a plateau actually emerges, however, also depends on the initial state.

The second condition depends on the initial state through the overlap products ${c_m^*}^{(0)}c_n^{(0)}$. Figures~\ref{fig:FIG03permut}(c) and \ref{fig:FIG03permut}(d) show these elements for two different $|\Psi_0\rangle$, with the red boxes delimiting the degenerate energy bands. For the initial state in Fig.~\ref{fig:FIG03permut}(c), the dominant intraband overlaps occur in the bands highlighted by the dark-blue boxes in Fig.~\ref{fig:FIG03permut}(b), where the off-diagonal matrix elements of $\hat C^z$ are negligible. As a result, the first term in Eq.~(\ref{eq:Obs_pre}) is therefore negligible, and no prethermal plateau develops. By contrast, for the initial state in
Fig.~\ref{fig:FIG03permut}(d), the dominant contributions occur within bands where $\hat C^z$ possesses large off-diagonal matrix elements [green
block in Fig.~\ref{fig:FIG03permut}(b)]. Both conditions are then satisfied, and a prethermal plateau emerges.

%These results demonstrate that, even when the Hamiltonian provides well-separated dynamical timescales, the emergence of a prethermal plateau depends jointly on the observable and the initial state.

%%%%%%%%%%%%%%%%%%%%%%%%%%%%%%%%%
%%%%%%%%%%%%%%%%%%%%%%%%%%%%%%%%%
%%%%%%%%%%%%%%%%%%%%%%%%%%%%%%%%%
{\em Lower bound on prethermal lifetimes.--} We now establish a general bound on the lifetime of prethermal plateaus. Consider an exact unitary evolution $\hat U(t)$ and a unitary prethermal evolution $\hat U_{\rm pre}(t)$. The departure of the exact dynamics from the prethermal evolution is quantified by the Loschmidt echo,
\begin{align}
\mathcal L(t) = \left|
\langle\Psi_0|
\hat U_{\rm pre}^\dagger(t)\hat U(t)
|\Psi_0\rangle
\right|^2.
\label{eq:LE}
\end{align}
For the Hamiltonian and pure initial states considered here, $\hat U(t)=e^{-i\hat H^{(\alpha)}t}$, $\hat U_{\rm pre}(t)=e^{-i\hat H^{(0)}t}$, and $\hat\rho_0=|\Psi_0\rangle\langle\Psi_0|$. However, the following derivation relies only on the unitarity of the evolutions, applying also to time-dependent Hamiltonians and admitting an extension to mixed states, as explained in the SM~\cite{noteSUPPL}.

To compare the dynamics of the Loschmidt echo with that of a generic bounded observable $\hat O$, we define the deviations,
\begin{align}
\chi_O(t) &= \left| \langle\hat O\rangle(t) - \langle\hat O\rangle_{\rm pre} \right|
= \left| {\rm Tr} \!\left[ \hat O \left( \hat\rho(t)-\hat\rho_{\rm pre}(t) \right) \right] \right|,
\\
\chi_{\mathcal L}(t) &=  |\mathcal{L}(t) - \mathcal{L}^{\text{pre}}(t) | = 1-\mathcal L(t).
\end{align}
Using the trace inequality 
$\left| {\rm Tr}[A B] \right| \le \|A\|_\infty \|B\|_1$,
where $\|\cdot\|_\infty$ denotes the operator norm (largest singular value) and $\|\cdot\|_1$ denotes the trace norm, 
together with 
\begin{align}
\| \hat\rho(t)-\hat\rho_{\rm pre}(t) \|_1 = 2\sqrt{1-\mathcal L(t)},
\end{align}
we obtain 
\begin{align}
\chi_O(t) \le 2 \|\hat O\|_\infty \sqrt{\chi_{\mathcal L}(t)}.
\label{eq:master_bound}
\end{align}

We define $t_O$ as the first instant at which $\chi_O(t)$ exceeds a prescribed threshold,
$t_O = \inf \{ t\ge0: \chi_O(t)\ge\delta_O \}$,
and define the corresponding Loschmidt-echo threshold, 
\begin{equation}
 \delta_{\mathcal L} = \delta_O^2/(4\|\hat O\|_\infty^2),
 \label{Eq:deltaO}
\end{equation}
through the saturation of Eq.~(\ref{eq:master_bound}), so that
$t_{\mathcal L} = \inf \{ t\ge0: \chi_{\mathcal L}(t) \ge \delta_{\mathcal L} 
\}$. 
The choice of $\delta_{\mathcal L}$ corresponds to the largest threshold compatible with Eq.~(\ref{eq:master_bound}). Whenever
$\chi_{\mathcal L}(t) < \delta_{\mathcal L}$, 
Eq.~(\ref{eq:master_bound}) implies that 
\begin{align}
\chi_O(t) < 2
\|\hat O\|_\infty
\sqrt{\delta_{\mathcal L}} =
\delta_O.
\end{align}
Therefore, $\chi_O(t)<\delta_O$ for all $t<t_{\mathcal L}$, and hence
\begin{align}
t_{\mathcal L}\le t_O.
\label{eq:tBound}
\end{align}
The prethermal lifetime of the Loschmidt echo therefore provides a lower bound on that of any bounded observable.

%%%%%%%%%%%%%%%%% FIGURE %%%%%%%%%%%
\begin{figure}[h]
    \begin{center}
        \includegraphics[width=0.9\columnwidth]{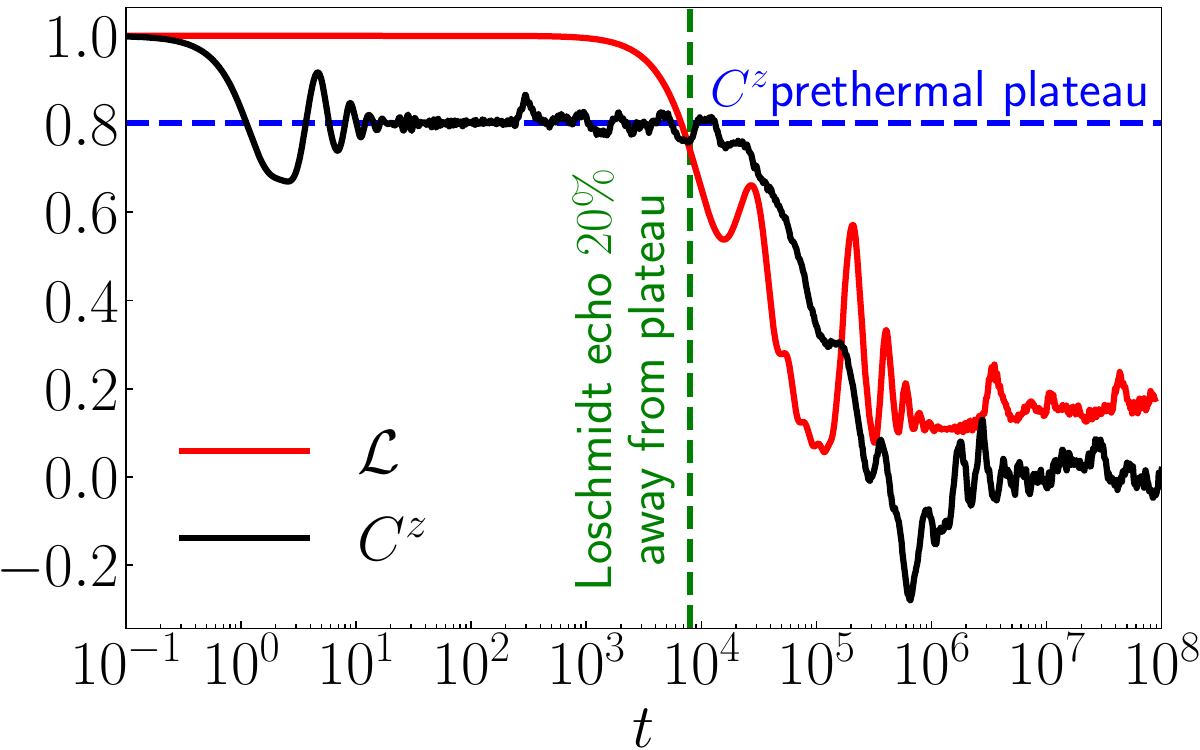}
        \caption{Loschmidt echo $\mathcal{L}$ and the observable $C^z$ for $L=12$ and $\alpha=10^{-4}$. Vertical dashed line marks the time at which the Loschmidt echo has departed from its prethermal regime by $\delta_{\mathcal L}\approx0.2$. At this time, $C^z (t)$ remains at its prethermal plateau, illustrating the bound $t_{\mathcal L} \le t_O$.
        }
        \label{fig:Loschmidt}
    \end{center}
\end{figure}

Figure~\ref{fig:Loschmidt} illustrates this bound for $C^z$. At the time when the Loschmidt echo (red curves) reaches a prescribed deviation $\delta_{\mathcal L}$, $C^z$ (black curve) remains at its prethermal plateau and is far from reaching the corresponding threshold $\delta_O$ in Eq.~(\ref{Eq:deltaO}), consistent with the bound in Eq.~(\ref{eq:tBound}).

%%%%%%%%%%%%%%%%%%%%%%%%%%%%%%%%%%%%%
%%%%%%%%%%%%%%%%%%%%%%%%%%%%%%%%%%%%%
%%%%%%%%%%%%%%%%%%%%%%%%%%%%%%%%%%%%%
{\em Conclusions.--} We developed a theoretical framework for prethermalization in quantum many-body systems with approximate permutation symmetry. The separation between fast interband and slow intraband dynamics creates a prethermal time window, but does not guarantee a prethermal plateau. The emergence of this plateau requires surviving intraband coherences, whose contribution is controlled by the observable and the initial state. This leads, in particular, to a symmetry-based selection rule that excludes prethermal plateaus for observables preserving full permutation symmetry. Thus, while the Hamiltonian sets the available dynamical timescales, the preparation and the probe determine whether these timescales become manifest as a prethermal plateau.

%Our results thus distinguish two fundamental aspects of prethermalization: the Hamiltonian sets the available dynamical timescales, while the preparation and the probe determine whether these timescales become manifest as a prethermal plateau.

We further proved that the Loschmidt echo provides a lower bound on the prethermal lifetime of any bounded observable. The bound is independent of the mechanism underlying prethermalization and requires only that the exact and prethermal dynamics are unitary, making it applicable to systems described by both time-independent and time-dependent Hamiltonians.

{\em Acknowledgements.--} C.L.S. and L.F.S are supported by UConn funds. S.K.P is supported in TIFR through a graduate fellowship from the Department of Atomic Energy (D.A.E),
Govt. of India.

%%%%%%%%%%%%%%%%%%%%%%%%%%%%%%%%%%%%%%%%%%%%%%%%%%%%%%%%%%%%%
%%%%%%%%%%%%%%%%%%% REFERENCES %%%%%%%%%%%%%%%%%%%%%%%%%%%%%%
%%%%%%%%%%%%%%%%%%%%%%%%%%%%%%%%%%%%%%%%%%%%%%%%%%%%%%%%%%%%%
\bibliography{biblio2025}

\begin{thebibliography}{48}%
\makeatletter
\providecommand \@ifxundefined [1]{%
 \@ifx{#1\undefined}
}%
\providecommand \@ifnum [1]{%
 \ifnum #1\expandafter \@firstoftwo
 \else \expandafter \@secondoftwo
 \fi
}%
\providecommand \@ifx [1]{%
 \ifx #1\expandafter \@firstoftwo
 \else \expandafter \@secondoftwo
 \fi
}%
\providecommand \natexlab [1]{#1}%
\providecommand \enquote  [1]{``#1''}%
\providecommand \bibnamefont  [1]{#1}%
\providecommand \bibfnamefont [1]{#1}%
\providecommand \citenamefont [1]{#1}%
\providecommand \href@noop [0]{\@secondoftwo}%
\providecommand \href [0]{\begingroup \@sanitize@url \@href}%
\providecommand \@href[1]{\@@startlink{#1}\@@href}%
\providecommand \@@href[1]{\endgroup#1\@@endlink}%
\providecommand \@sanitize@url [0]{\catcode `\\12\catcode `\$12\catcode `\&12\catcode `\#12\catcode `\^12\catcode `\_12\catcode `\%12\relax}%
\providecommand \@@startlink[1]{}%
\providecommand \@@endlink[0]{}%
\providecommand \url  [0]{\begingroup\@sanitize@url \@url }%
\providecommand \@url [1]{\endgroup\@href {#1}{\urlprefix }}%
\providecommand \urlprefix  [0]{URL }%
\providecommand \Eprint [0]{\href }%
\providecommand \doibase [0]{https://doi.org/}%
\providecommand \selectlanguage [0]{\@gobble}%
\providecommand \bibinfo  [0]{\@secondoftwo}%
\providecommand \bibfield  [0]{\@secondoftwo}%
\providecommand \translation [1]{[#1]}%
\providecommand \BibitemOpen [0]{}%
\providecommand \bibitemStop [0]{}%
\providecommand \bibitemNoStop [0]{.\EOS\space}%
\providecommand \EOS [0]{\spacefactor3000\relax}%
\providecommand \BibitemShut  [1]{\csname bibitem#1\endcsname}%
\let\auto@bib@innerbib\@empty
%</preamble>
\bibitem [{\citenamefont {Barnett}\ \emph {et~al.}(2011)\citenamefont {Barnett}, \citenamefont {Polkovnikov},\ and\ \citenamefont {Vengalattore}}]{Barnett2011}%
  \BibitemOpen
  \bibfield  {author} {\bibinfo {author} {\bibfnamefont {R.}~\bibnamefont {Barnett}}, \bibinfo {author} {\bibfnamefont {A.}~\bibnamefont {Polkovnikov}},\ and\ \bibinfo {author} {\bibfnamefont {M.}~\bibnamefont {Vengalattore}},\ }\bibfield  {title} {\bibinfo {title} {Prethermalization in quenched spinor condensates},\ }\href {https://doi.org/10.1103/PhysRevA.84.023606} {\bibfield  {journal} {\bibinfo  {journal} {Phys. Rev. A}\ }\textbf {\bibinfo {volume} {84}},\ \bibinfo {pages} {023606} (\bibinfo {year} {2011})}\BibitemShut {NoStop}%
\bibitem [{\citenamefont {Kollar}\ \emph {et~al.}(2011)\citenamefont {Kollar}, \citenamefont {Wolf},\ and\ \citenamefont {Eckstein}}]{Kollar2011}%
  \BibitemOpen
  \bibfield  {author} {\bibinfo {author} {\bibfnamefont {M.}~\bibnamefont {Kollar}}, \bibinfo {author} {\bibfnamefont {F.~A.}\ \bibnamefont {Wolf}},\ and\ \bibinfo {author} {\bibfnamefont {M.}~\bibnamefont {Eckstein}},\ }\bibfield  {title} {\bibinfo {title} {Generalized gibbs ensemble prediction of prethermalization plateaus and their relation to nonthermal steady states in integrable systems},\ }\href {https://doi.org/10.1103/PhysRevB.84.054304} {\bibfield  {journal} {\bibinfo  {journal} {Phys. Rev. B}\ }\textbf {\bibinfo {volume} {84}},\ \bibinfo {pages} {054304} (\bibinfo {year} {2011})}\BibitemShut {NoStop}%
\bibitem [{\citenamefont {Babadi}\ \emph {et~al.}(2015)\citenamefont {Babadi}, \citenamefont {Demler},\ and\ \citenamefont {Knap}}]{Babadi2015}%
  \BibitemOpen
  \bibfield  {author} {\bibinfo {author} {\bibfnamefont {M.}~\bibnamefont {Babadi}}, \bibinfo {author} {\bibfnamefont {E.}~\bibnamefont {Demler}},\ and\ \bibinfo {author} {\bibfnamefont {M.}~\bibnamefont {Knap}},\ }\bibfield  {title} {\bibinfo {title} {Far-from-equilibrium field theory of many-body quantum spin systems: Prethermalization and relaxation of spin spiral states in three dimensions},\ }\href {https://doi.org/10.1103/PhysRevX.5.041005} {\bibfield  {journal} {\bibinfo  {journal} {Phys. Rev. X}\ }\textbf {\bibinfo {volume} {5}},\ \bibinfo {pages} {041005} (\bibinfo {year} {2015})}\BibitemShut {NoStop}%
\bibitem [{\citenamefont {Alba}\ and\ \citenamefont {Fagotti}(2017)}]{Alba2017}%
  \BibitemOpen
  \bibfield  {author} {\bibinfo {author} {\bibfnamefont {V.}~\bibnamefont {Alba}}\ and\ \bibinfo {author} {\bibfnamefont {M.}~\bibnamefont {Fagotti}},\ }\bibfield  {title} {\bibinfo {title} {Prethermalization at low temperature: The scent of long-range order},\ }\href {https://doi.org/10.1103/PhysRevLett.119.010601} {\bibfield  {journal} {\bibinfo  {journal} {Phys. Rev. Lett.}\ }\textbf {\bibinfo {volume} {119}},\ \bibinfo {pages} {010601} (\bibinfo {year} {2017})}\BibitemShut {NoStop}%
\bibitem [{\citenamefont {Abanin}\ \emph {et~al.}(2017)\citenamefont {Abanin}, \citenamefont {De~Roeck}, \citenamefont {Ho},\ and\ \citenamefont {Huveneers}}]{Abanin2017}%
  \BibitemOpen
  \bibfield  {author} {\bibinfo {author} {\bibfnamefont {D.}~\bibnamefont {Abanin}}, \bibinfo {author} {\bibfnamefont {W.}~\bibnamefont {De~Roeck}}, \bibinfo {author} {\bibfnamefont {W.~W.}\ \bibnamefont {Ho}},\ and\ \bibinfo {author} {\bibfnamefont {F.}~\bibnamefont {Huveneers}},\ }\bibfield  {title} {\bibinfo {title} {A rigorous theory of many-body prethermalization for periodically driven and closed quantum systems},\ }\href {https://doi.org/10.1007/s00220-017-2930-x} {\bibfield  {journal} {\bibinfo  {journal} {Comm. Math. Phys.}\ }\textbf {\bibinfo {volume} {354}},\ \bibinfo {pages} {809} (\bibinfo {year} {2017})}\BibitemShut {NoStop}%
\bibitem [{\citenamefont {Mori}\ \emph {et~al.}(2018)\citenamefont {Mori}, \citenamefont {Ikeda}, \citenamefont {Kaminishi},\ and\ \citenamefont {Ueda}}]{Mori2018}%
  \BibitemOpen
  \bibfield  {author} {\bibinfo {author} {\bibfnamefont {T.}~\bibnamefont {Mori}}, \bibinfo {author} {\bibfnamefont {T.~N.}\ \bibnamefont {Ikeda}}, \bibinfo {author} {\bibfnamefont {E.}~\bibnamefont {Kaminishi}},\ and\ \bibinfo {author} {\bibfnamefont {M.}~\bibnamefont {Ueda}},\ }\bibfield  {title} {\bibinfo {title} {Thermalization and prethermalization in isolated quantum systems: A theoretical overview},\ }\href {https://doi.org/10.1088/1361-6455/aabcdf} {\bibfield  {journal} {\bibinfo  {journal} {J. Phys. B}\ }\textbf {\bibinfo {volume} {51}},\ \bibinfo {pages} {112001} (\bibinfo {year} {2018})}\BibitemShut {NoStop}%
\bibitem [{\citenamefont {Reimann}\ and\ \citenamefont {Dabelow}(2019)}]{Reimann2019_01}%
  \BibitemOpen
  \bibfield  {author} {\bibinfo {author} {\bibfnamefont {P.}~\bibnamefont {Reimann}}\ and\ \bibinfo {author} {\bibfnamefont {L.}~\bibnamefont {Dabelow}},\ }\bibfield  {title} {\bibinfo {title} {Typicality of prethermalization},\ }\href {https://doi.org/10.1103/PhysRevLett.122.080603} {\bibfield  {journal} {\bibinfo  {journal} {Phys. Rev. Lett.}\ }\textbf {\bibinfo {volume} {122}},\ \bibinfo {pages} {080603} (\bibinfo {year} {2019})}\BibitemShut {NoStop}%
\bibitem [{\citenamefont {Yin}\ and\ \citenamefont {Lucas}(2023)}]{Yin2023}%
  \BibitemOpen
  \bibfield  {author} {\bibinfo {author} {\bibfnamefont {C.}~\bibnamefont {Yin}}\ and\ \bibinfo {author} {\bibfnamefont {A.}~\bibnamefont {Lucas}},\ }\bibfield  {title} {\bibinfo {title} {Prethermalization and the local robustness of gapped systems},\ }\href {https://doi.org/10.1103/PhysRevLett.131.050402} {\bibfield  {journal} {\bibinfo  {journal} {Phys. Rev. Lett.}\ }\textbf {\bibinfo {volume} {131}},\ \bibinfo {pages} {050402} (\bibinfo {year} {2023})}\BibitemShut {NoStop}%
\bibitem [{\citenamefont {Gallone}(2026)}]{Gallone2026}%
  \BibitemOpen
  \bibfield  {author} {\bibinfo {author} {\bibfnamefont {M.}~\bibnamefont {Gallone}},\ }\href {https://arxiv.org/abs/2604.13781} {\bibinfo {title} {On exponentially long prethermalization timescales in isolated quantum systems}} (\bibinfo {year} {2026}),\ \Eprint {https://arxiv.org/abs/2604.13781} {arXiv:2604.13781 [math-ph]} \BibitemShut {NoStop}%
\bibitem [{\citenamefont {Bertini}\ \emph {et~al.}(2015)\citenamefont {Bertini}, \citenamefont {Essler}, \citenamefont {Groha},\ and\ \citenamefont {Robinson}}]{Bertini2015}%
  \BibitemOpen
  \bibfield  {author} {\bibinfo {author} {\bibfnamefont {B.}~\bibnamefont {Bertini}}, \bibinfo {author} {\bibfnamefont {F.~H.~L.}\ \bibnamefont {Essler}}, \bibinfo {author} {\bibfnamefont {S.}~\bibnamefont {Groha}},\ and\ \bibinfo {author} {\bibfnamefont {N.~J.}\ \bibnamefont {Robinson}},\ }\bibfield  {title} {\bibinfo {title} {Prethermalization and thermalization in models with weak integrability breaking},\ }\href {https://doi.org/10.1103/PhysRevLett.115.180601} {\bibfield  {journal} {\bibinfo  {journal} {Phys. Rev. Lett.}\ }\textbf {\bibinfo {volume} {115}},\ \bibinfo {pages} {180601} (\bibinfo {year} {2015})}\BibitemShut {NoStop}%
\bibitem [{\citenamefont {Else}\ \emph {et~al.}(2017)\citenamefont {Else}, \citenamefont {Fendley}, \citenamefont {Kemp},\ and\ \citenamefont {Nayak}}]{Else2017}%
  \BibitemOpen
  \bibfield  {author} {\bibinfo {author} {\bibfnamefont {D.~V.}\ \bibnamefont {Else}}, \bibinfo {author} {\bibfnamefont {P.}~\bibnamefont {Fendley}}, \bibinfo {author} {\bibfnamefont {J.}~\bibnamefont {Kemp}},\ and\ \bibinfo {author} {\bibfnamefont {C.}~\bibnamefont {Nayak}},\ }\bibfield  {title} {\bibinfo {title} {Prethermal strong zero modes and topological qubits},\ }\href {https://doi.org/10.1103/PhysRevX.7.041062} {\bibfield  {journal} {\bibinfo  {journal} {Phys. Rev. X}\ }\textbf {\bibinfo {volume} {7}},\ \bibinfo {pages} {041062} (\bibinfo {year} {2017})}\BibitemShut {NoStop}%
\bibitem [{\citenamefont {Mallayya}\ \emph {et~al.}(2019)\citenamefont {Mallayya}, \citenamefont {Rigol},\ and\ \citenamefont {De~Roeck}}]{Mallayya2019}%
  \BibitemOpen
  \bibfield  {author} {\bibinfo {author} {\bibfnamefont {K.}~\bibnamefont {Mallayya}}, \bibinfo {author} {\bibfnamefont {M.}~\bibnamefont {Rigol}},\ and\ \bibinfo {author} {\bibfnamefont {W.}~\bibnamefont {De~Roeck}},\ }\bibfield  {title} {\bibinfo {title} {Prethermalization and thermalization in isolated quantum systems},\ }\href {https://doi.org/10.1103/PhysRevX.9.021027} {\bibfield  {journal} {\bibinfo  {journal} {Phys. Rev. X}\ }\textbf {\bibinfo {volume} {9}},\ \bibinfo {pages} {021027} (\bibinfo {year} {2019})}\BibitemShut {NoStop}%
\bibitem [{\citenamefont {Mallayya}\ and\ \citenamefont {Rigol}(2021)}]{Mallayya2021}%
  \BibitemOpen
  \bibfield  {author} {\bibinfo {author} {\bibfnamefont {K.}~\bibnamefont {Mallayya}}\ and\ \bibinfo {author} {\bibfnamefont {M.}~\bibnamefont {Rigol}},\ }\bibfield  {title} {\bibinfo {title} {Prethermalization, thermalization, and fermi's golden rule in quantum many-body systems},\ }\href {https://doi.org/10.1103/PhysRevB.104.184302} {\bibfield  {journal} {\bibinfo  {journal} {Phys. Rev. B}\ }\textbf {\bibinfo {volume} {104}},\ \bibinfo {pages} {184302} (\bibinfo {year} {2021})}\BibitemShut {NoStop}%
\bibitem [{\citenamefont {Kuwahara}\ \emph {et~al.}(2016)\citenamefont {Kuwahara}, \citenamefont {Mori},\ and\ \citenamefont {Saito}}]{Kuwahara2016}%
  \BibitemOpen
  \bibfield  {author} {\bibinfo {author} {\bibfnamefont {T.}~\bibnamefont {Kuwahara}}, \bibinfo {author} {\bibfnamefont {T.}~\bibnamefont {Mori}},\ and\ \bibinfo {author} {\bibfnamefont {K.}~\bibnamefont {Saito}},\ }\bibfield  {title} {\bibinfo {title} {Floquet--magnus theory and generic transient dynamics in periodically driven many-body quantum systems},\ }\href {https://doi.org/10.1016/j.aop.2016.01.012} {\bibfield  {journal} {\bibinfo  {journal} {Annals of Physics}\ }\textbf {\bibinfo {volume} {367}},\ \bibinfo {pages} {96} (\bibinfo {year} {2016})}\BibitemShut {NoStop}%
\bibitem [{\citenamefont {Ho}\ \emph {et~al.}(2018)\citenamefont {Ho}, \citenamefont {Protopopov},\ and\ \citenamefont {Abanin}}]{Ho2018}%
  \BibitemOpen
  \bibfield  {author} {\bibinfo {author} {\bibfnamefont {W.~W.}\ \bibnamefont {Ho}}, \bibinfo {author} {\bibfnamefont {I.}~\bibnamefont {Protopopov}},\ and\ \bibinfo {author} {\bibfnamefont {D.~A.}\ \bibnamefont {Abanin}},\ }\bibfield  {title} {\bibinfo {title} {Bounds on energy absorption and prethermalization in quantum systems with long-range interactions},\ }\href {https://doi.org/10.1103/PhysRevLett.120.200601} {\bibfield  {journal} {\bibinfo  {journal} {Phys. Rev. Lett.}\ }\textbf {\bibinfo {volume} {120}},\ \bibinfo {pages} {200601} (\bibinfo {year} {2018})}\BibitemShut {NoStop}%
\bibitem [{\citenamefont {Dumitrescu}\ \emph {et~al.}(2018)\citenamefont {Dumitrescu}, \citenamefont {Vasseur},\ and\ \citenamefont {Potter}}]{Dumitrescu2018}%
  \BibitemOpen
  \bibfield  {author} {\bibinfo {author} {\bibfnamefont {P.~T.}\ \bibnamefont {Dumitrescu}}, \bibinfo {author} {\bibfnamefont {R.}~\bibnamefont {Vasseur}},\ and\ \bibinfo {author} {\bibfnamefont {A.~C.}\ \bibnamefont {Potter}},\ }\bibfield  {title} {\bibinfo {title} {Logarithmically slow relaxation in quasiperiodically driven random spin chains},\ }\href {https://doi.org/10.1103/PhysRevLett.120.070602} {\bibfield  {journal} {\bibinfo  {journal} {Phys. Rev. Lett.}\ }\textbf {\bibinfo {volume} {120}},\ \bibinfo {pages} {070602} (\bibinfo {year} {2018})}\BibitemShut {NoStop}%
\bibitem [{\citenamefont {Machado}\ \emph {et~al.}(2019)\citenamefont {Machado}, \citenamefont {Kahanamoku-Meyer}, \citenamefont {Else}, \citenamefont {Nayak},\ and\ \citenamefont {Yao}}]{Machado2019}%
  \BibitemOpen
  \bibfield  {author} {\bibinfo {author} {\bibfnamefont {F.}~\bibnamefont {Machado}}, \bibinfo {author} {\bibfnamefont {G.~D.}\ \bibnamefont {Kahanamoku-Meyer}}, \bibinfo {author} {\bibfnamefont {D.~V.}\ \bibnamefont {Else}}, \bibinfo {author} {\bibfnamefont {C.}~\bibnamefont {Nayak}},\ and\ \bibinfo {author} {\bibfnamefont {N.~Y.}\ \bibnamefont {Yao}},\ }\bibfield  {title} {\bibinfo {title} {Exponentially slow heating in short and long-range interacting floquet systems},\ }\href {https://doi.org/10.1103/PhysRevResearch.1.033202} {\bibfield  {journal} {\bibinfo  {journal} {Phys. Rev. Res.}\ }\textbf {\bibinfo {volume} {1}},\ \bibinfo {pages} {033202} (\bibinfo {year} {2019})}\BibitemShut {NoStop}%
\bibitem [{\citenamefont {Machado}\ \emph {et~al.}(2020)\citenamefont {Machado}, \citenamefont {Else}, \citenamefont {Kahanamoku-Meyer}, \citenamefont {Nayak},\ and\ \citenamefont {Yao}}]{Machado2020}%
  \BibitemOpen
  \bibfield  {author} {\bibinfo {author} {\bibfnamefont {F.}~\bibnamefont {Machado}}, \bibinfo {author} {\bibfnamefont {D.~V.}\ \bibnamefont {Else}}, \bibinfo {author} {\bibfnamefont {G.~D.}\ \bibnamefont {Kahanamoku-Meyer}}, \bibinfo {author} {\bibfnamefont {C.}~\bibnamefont {Nayak}},\ and\ \bibinfo {author} {\bibfnamefont {N.~Y.}\ \bibnamefont {Yao}},\ }\bibfield  {title} {\bibinfo {title} {Long-range prethermal phases of nonequilibrium matter},\ }\href {https://doi.org/10.1103/PhysRevX.10.011043} {\bibfield  {journal} {\bibinfo  {journal} {Phys. Rev. X}\ }\textbf {\bibinfo {volume} {10}},\ \bibinfo {pages} {011043} (\bibinfo {year} {2020})}\BibitemShut {NoStop}%
\bibitem [{\citenamefont {Mori}\ \emph {et~al.}(2021)\citenamefont {Mori}, \citenamefont {Zhao}, \citenamefont {Mintert}, \citenamefont {Knolle},\ and\ \citenamefont {Moessner}}]{Mori2021}%
  \BibitemOpen
  \bibfield  {author} {\bibinfo {author} {\bibfnamefont {T.}~\bibnamefont {Mori}}, \bibinfo {author} {\bibfnamefont {H.}~\bibnamefont {Zhao}}, \bibinfo {author} {\bibfnamefont {F.}~\bibnamefont {Mintert}}, \bibinfo {author} {\bibfnamefont {J.}~\bibnamefont {Knolle}},\ and\ \bibinfo {author} {\bibfnamefont {R.}~\bibnamefont {Moessner}},\ }\bibfield  {title} {\bibinfo {title} {Rigorous bounds on the heating rate in thue-morse quasiperiodically and randomly driven quantum many-body systems},\ }\href {https://doi.org/10.1103/PhysRevLett.127.050602} {\bibfield  {journal} {\bibinfo  {journal} {Phys. Rev. Lett.}\ }\textbf {\bibinfo {volume} {127}},\ \bibinfo {pages} {050602} (\bibinfo {year} {2021})}\BibitemShut {NoStop}%
\bibitem [{\citenamefont {Santos}(2021)}]{Santos2021}%
  \BibitemOpen
  \bibfield  {author} {\bibinfo {author} {\bibfnamefont {L.~F.}\ \bibnamefont {Santos}},\ }\bibfield  {title} {\bibinfo {title} {The quick drive to pseudo-equilibrium},\ }\href {https://doi.org/10.1038/s41567-020-01117-8} {\bibfield  {journal} {\bibinfo  {journal} {Nat. Phys.}\ }\textbf {\bibinfo {volume} {17}},\ \bibinfo {pages} {429} (\bibinfo {year} {2021})}\BibitemShut {NoStop}%
\bibitem [{\citenamefont {Bhakuni}\ \emph {et~al.}(2021)\citenamefont {Bhakuni}, \citenamefont {Santos},\ and\ \citenamefont {Lev}}]{Bhakuni2021}%
  \BibitemOpen
  \bibfield  {author} {\bibinfo {author} {\bibfnamefont {D.~S.}\ \bibnamefont {Bhakuni}}, \bibinfo {author} {\bibfnamefont {L.~F.}\ \bibnamefont {Santos}},\ and\ \bibinfo {author} {\bibfnamefont {Y.~B.}\ \bibnamefont {Lev}},\ }\bibfield  {title} {\bibinfo {title} {Suppression of heating by long-range interactions in periodically driven spin chains},\ }\href {https://doi.org/10.1103/PhysRevB.104.L140301} {\bibfield  {journal} {\bibinfo  {journal} {Phys. Rev. B}\ }\textbf {\bibinfo {volume} {104}},\ \bibinfo {pages} {L140301} (\bibinfo {year} {2021})}\BibitemShut {NoStop}%
\bibitem [{\citenamefont {Das}\ \emph {et~al.}(2023)\citenamefont {Das}, \citenamefont {Bhakuni}, \citenamefont {Santos},\ and\ \citenamefont {Sharma}}]{Das2023}%
  \BibitemOpen
  \bibfield  {author} {\bibinfo {author} {\bibfnamefont {P.}~\bibnamefont {Das}}, \bibinfo {author} {\bibfnamefont {D.~S.}\ \bibnamefont {Bhakuni}}, \bibinfo {author} {\bibfnamefont {L.~F.}\ \bibnamefont {Santos}},\ and\ \bibinfo {author} {\bibfnamefont {A.}~\bibnamefont {Sharma}},\ }\bibfield  {title} {\bibinfo {title} {Periodically and quasiperiodically driven anisotropic dicke model},\ }\href {https://doi.org/10.1103/PhysRevA.108.063716} {\bibfield  {journal} {\bibinfo  {journal} {Phys. Rev. A}\ }\textbf {\bibinfo {volume} {108}},\ \bibinfo {pages} {063716} (\bibinfo {year} {2023})}\BibitemShut {NoStop}%
\bibitem [{\citenamefont {O'Dea}\ \emph {et~al.}(2024)\citenamefont {O'Dea}, \citenamefont {Burnell}, \citenamefont {Chandran},\ and\ \citenamefont {Khemani}}]{Dea2024}%
  \BibitemOpen
  \bibfield  {author} {\bibinfo {author} {\bibfnamefont {N.}~\bibnamefont {O'Dea}}, \bibinfo {author} {\bibfnamefont {F.}~\bibnamefont {Burnell}}, \bibinfo {author} {\bibfnamefont {A.}~\bibnamefont {Chandran}},\ and\ \bibinfo {author} {\bibfnamefont {V.}~\bibnamefont {Khemani}},\ }\bibfield  {title} {\bibinfo {title} {Prethermal stability of eigenstates under high frequency floquet driving},\ }\href {https://doi.org/10.1103/PhysRevLett.132.100401} {\bibfield  {journal} {\bibinfo  {journal} {Phys. Rev. Lett.}\ }\textbf {\bibinfo {volume} {132}},\ \bibinfo {pages} {100401} (\bibinfo {year} {2024})}\BibitemShut {NoStop}%
\bibitem [{\citenamefont {Tiwari}\ \emph {et~al.}(2025)\citenamefont {Tiwari}, \citenamefont {Bhakuni},\ and\ \citenamefont {Sharma}}]{Tiwari2025}%
  \BibitemOpen
  \bibfield  {author} {\bibinfo {author} {\bibfnamefont {V.}~\bibnamefont {Tiwari}}, \bibinfo {author} {\bibfnamefont {D.~S.}\ \bibnamefont {Bhakuni}},\ and\ \bibinfo {author} {\bibfnamefont {A.}~\bibnamefont {Sharma}},\ }\bibfield  {title} {\bibinfo {title} {Periodically and aperiodically thue-morse driven long-range systems: From dynamical localization to slow dynamics},\ }\href {https://doi.org/10.1103/PhysRevB.111.205109} {\bibfield  {journal} {\bibinfo  {journal} {Phys. Rev. B}\ }\textbf {\bibinfo {volume} {111}},\ \bibinfo {pages} {205109} (\bibinfo {year} {2025})}\BibitemShut {NoStop}%
\bibitem [{\citenamefont {Kastner}(2010)}]{Kastner2010}%
  \BibitemOpen
  \bibfield  {author} {\bibinfo {author} {\bibfnamefont {M.}~\bibnamefont {Kastner}},\ }\bibfield  {title} {\bibinfo {title} {Nonequivalence of ensembles for long-range quantum spin systems in optical lattices},\ }\href {https://doi.org/10.1103/PhysRevLett.104.240403} {\bibfield  {journal} {\bibinfo  {journal} {Phys. Rev. Lett.}\ }\textbf {\bibinfo {volume} {104}},\ \bibinfo {pages} {240403} (\bibinfo {year} {2010})}\BibitemShut {NoStop}%
\bibitem [{\citenamefont {Sch\"utz}\ and\ \citenamefont {Morigi}(2014)}]{Schutz2014}%
  \BibitemOpen
  \bibfield  {author} {\bibinfo {author} {\bibfnamefont {S.}~\bibnamefont {Sch\"utz}}\ and\ \bibinfo {author} {\bibfnamefont {G.}~\bibnamefont {Morigi}},\ }\bibfield  {title} {\bibinfo {title} {Prethermalization of atoms due to photon-mediated long-range interactions},\ }\href {https://doi.org/10.1103/PhysRevLett.113.203002} {\bibfield  {journal} {\bibinfo  {journal} {Phys. Rev. Lett.}\ }\textbf {\bibinfo {volume} {113}},\ \bibinfo {pages} {203002} (\bibinfo {year} {2014})}\BibitemShut {NoStop}%
\bibitem [{\citenamefont {Sch\"utz}\ \emph {et~al.}(2016)\citenamefont {Sch\"utz}, \citenamefont {J\"ager},\ and\ \citenamefont {Morigi}}]{Schutz2016}%
  \BibitemOpen
  \bibfield  {author} {\bibinfo {author} {\bibfnamefont {S.}~\bibnamefont {Sch\"utz}}, \bibinfo {author} {\bibfnamefont {S.~B.}\ \bibnamefont {J\"ager}},\ and\ \bibinfo {author} {\bibfnamefont {G.}~\bibnamefont {Morigi}},\ }\bibfield  {title} {\bibinfo {title} {Dissipation-assisted prethermalization in long-range interacting atomic ensembles},\ }\href {https://doi.org/10.1103/PhysRevLett.117.083001} {\bibfield  {journal} {\bibinfo  {journal} {Phys. Rev. Lett.}\ }\textbf {\bibinfo {volume} {117}},\ \bibinfo {pages} {083001} (\bibinfo {year} {2016})}\BibitemShut {NoStop}%
\bibitem [{\citenamefont {Mori}(2019)}]{Mori2019}%
  \BibitemOpen
  \bibfield  {author} {\bibinfo {author} {\bibfnamefont {T.}~\bibnamefont {Mori}},\ }\bibfield  {title} {\bibinfo {title} {Prethermalization in the transverse-field ising chain with long-range interactions},\ }\href {https://doi.org/10.1088/1751-8121/aaf9db} {\bibfield  {journal} {\bibinfo  {journal} {J. Phys. A}\ }\textbf {\bibinfo {volume} {52}},\ \bibinfo {pages} {054001} (\bibinfo {year} {2019})}\BibitemShut {NoStop}%
\bibitem [{\citenamefont {Defenu}\ \emph {et~al.}(2024)\citenamefont {Defenu}, \citenamefont {Mukamel},\ and\ \citenamefont {Ruffo}}]{Defenu2024_1}%
  \BibitemOpen
  \bibfield  {author} {\bibinfo {author} {\bibfnamefont {N.}~\bibnamefont {Defenu}}, \bibinfo {author} {\bibfnamefont {D.}~\bibnamefont {Mukamel}},\ and\ \bibinfo {author} {\bibfnamefont {S.}~\bibnamefont {Ruffo}},\ }\bibfield  {title} {\bibinfo {title} {Ensemble inequivalence in long-range quantum systems},\ }\href {https://doi.org/10.1103/PhysRevLett.133.050403} {\bibfield  {journal} {\bibinfo  {journal} {Phys. Rev. Lett.}\ }\textbf {\bibinfo {volume} {133}},\ \bibinfo {pages} {050403} (\bibinfo {year} {2024})}\BibitemShut {NoStop}%
\bibitem [{\citenamefont {Arrufat-Vicente}\ \emph {et~al.}(2026)\citenamefont {Arrufat-Vicente}, \citenamefont {Mukamel}, \citenamefont {Ruffo},\ and\ \citenamefont {Defenu}}]{ArrufatVicente2025}%
  \BibitemOpen
  \bibfield  {author} {\bibinfo {author} {\bibfnamefont {D.}~\bibnamefont {Arrufat-Vicente}}, \bibinfo {author} {\bibfnamefont {D.}~\bibnamefont {Mukamel}}, \bibinfo {author} {\bibfnamefont {S.}~\bibnamefont {Ruffo}},\ and\ \bibinfo {author} {\bibfnamefont {N.}~\bibnamefont {Defenu}},\ }\bibfield  {title} {\bibinfo {title} {Ensemble inequivalence in long-range quantum spin systems},\ }\bibfield  {journal} {\bibinfo  {journal} {Phys. Rev. Res.}\ }\href {https://doi.org/10.1103/hmhz-4g46} {10.1103/hmhz-4g46} (\bibinfo {year} {2026})\BibitemShut {NoStop}%
\bibitem [{\citenamefont {C}\ and\ \citenamefont {Divakaran}(2026)}]{Manju2026}%
  \BibitemOpen
  \bibfield  {author} {\bibinfo {author} {\bibfnamefont {M.}~\bibnamefont {C}}\ and\ \bibinfo {author} {\bibfnamefont {U.}~\bibnamefont {Divakaran}},\ }\bibfield  {title} {\bibinfo {title} {Floquet thermalization by power-law induced permutation symmetry breaking},\ }\href {https://doi.org/10.1103/5dwc-mqgp} {\bibfield  {journal} {\bibinfo  {journal} {Phys. Rev. E}\ }\textbf {\bibinfo {volume} {113}},\ \bibinfo {pages} {044209} (\bibinfo {year} {2026})}\BibitemShut {NoStop}%
\bibitem [{\citenamefont {Halimeh}\ and\ \citenamefont {Hauke}(2020)}]{Halimeh2020}%
  \BibitemOpen
  \bibfield  {author} {\bibinfo {author} {\bibfnamefont {J.~C.}\ \bibnamefont {Halimeh}}\ and\ \bibinfo {author} {\bibfnamefont {P.}~\bibnamefont {Hauke}},\ }\href {https://arxiv.org/abs/2004.07248} {\bibinfo {title} {Staircase prethermalization and constrained dynamics in lattice gauge theories}} (\bibinfo {year} {2020}),\ \Eprint {https://arxiv.org/abs/2004.07248} {arXiv:2004.07248 [cond-mat.quant-gas]} \BibitemShut {NoStop}%
\bibitem [{\citenamefont {Hayata}\ \emph {et~al.}(2024)\citenamefont {Hayata}, \citenamefont {Seki},\ and\ \citenamefont {Yamamoto}}]{Hayata2024}%
  \BibitemOpen
  \bibfield  {author} {\bibinfo {author} {\bibfnamefont {T.}~\bibnamefont {Hayata}}, \bibinfo {author} {\bibfnamefont {K.}~\bibnamefont {Seki}},\ and\ \bibinfo {author} {\bibfnamefont {A.}~\bibnamefont {Yamamoto}},\ }\bibfield  {title} {\bibinfo {title} {Floquet prethermalization of ${Z}_{2}$ lattice gauge theory on superconducting qubits},\ }\href {https://doi.org/10.1103/PhysRevD.110.114503} {\bibfield  {journal} {\bibinfo  {journal} {Phys. Rev. D}\ }\textbf {\bibinfo {volume} {110}},\ \bibinfo {pages} {114503} (\bibinfo {year} {2024})}\BibitemShut {NoStop}%
\bibitem [{\citenamefont {Tang}\ \emph {et~al.}(2018)\citenamefont {Tang}, \citenamefont {Kao}, \citenamefont {Li}, \citenamefont {Seo}, \citenamefont {Mallayya}, \citenamefont {Rigol}, \citenamefont {Gopalakrishnan},\ and\ \citenamefont {Lev}}]{Tang2018}%
  \BibitemOpen
  \bibfield  {author} {\bibinfo {author} {\bibfnamefont {Y.}~\bibnamefont {Tang}}, \bibinfo {author} {\bibfnamefont {W.}~\bibnamefont {Kao}}, \bibinfo {author} {\bibfnamefont {K.-Y.}\ \bibnamefont {Li}}, \bibinfo {author} {\bibfnamefont {S.}~\bibnamefont {Seo}}, \bibinfo {author} {\bibfnamefont {K.}~\bibnamefont {Mallayya}}, \bibinfo {author} {\bibfnamefont {M.}~\bibnamefont {Rigol}}, \bibinfo {author} {\bibfnamefont {S.}~\bibnamefont {Gopalakrishnan}},\ and\ \bibinfo {author} {\bibfnamefont {B.~L.}\ \bibnamefont {Lev}},\ }\bibfield  {title} {\bibinfo {title} {Thermalization near integrability in a dipolar quantum {N}ewton's cradle},\ }\href {https://doi.org/10.1103/PhysRevX.8.021030} {\bibfield  {journal} {\bibinfo  {journal} {Phys. Rev. X}\ }\textbf {\bibinfo {volume} {8}},\ \bibinfo {pages} {021030} (\bibinfo {year} {2018})}\BibitemShut {NoStop}%
\bibitem [{\citenamefont {Neyenhuis}\ \emph {et~al.}(2017)\citenamefont {Neyenhuis}, \citenamefont {Zhang}, \citenamefont {Hess}, \citenamefont {Smith}, \citenamefont {Lee}, \citenamefont {Richerme}, \citenamefont {Gong}, \citenamefont {Gorshkov},\ and\ \citenamefont {Monroe}}]{Neyenhuise2017}%
  \BibitemOpen
  \bibfield  {author} {\bibinfo {author} {\bibfnamefont {B.}~\bibnamefont {Neyenhuis}}, \bibinfo {author} {\bibfnamefont {J.}~\bibnamefont {Zhang}}, \bibinfo {author} {\bibfnamefont {P.~W.}\ \bibnamefont {Hess}}, \bibinfo {author} {\bibfnamefont {J.}~\bibnamefont {Smith}}, \bibinfo {author} {\bibfnamefont {A.~C.}\ \bibnamefont {Lee}}, \bibinfo {author} {\bibfnamefont {P.}~\bibnamefont {Richerme}}, \bibinfo {author} {\bibfnamefont {Z.-X.}\ \bibnamefont {Gong}}, \bibinfo {author} {\bibfnamefont {A.~V.}\ \bibnamefont {Gorshkov}},\ and\ \bibinfo {author} {\bibfnamefont {C.}~\bibnamefont {Monroe}},\ }\bibfield  {title} {\bibinfo {title} {Observation of prethermalization in long-range interacting spin chains},\ }\href {https://doi.org/10.1126/sciadv.1700672} {\bibfield  {journal} {\bibinfo  {journal} {Science Advances}\ }\textbf {\bibinfo {volume} {3}},\ \bibinfo {pages} {e1700672} (\bibinfo {year} {2017})}\BibitemShut {NoStop}%
\bibitem [{\citenamefont {Wei}\ \emph {et~al.}(2019)\citenamefont {Wei}, \citenamefont {Peng}, \citenamefont {Shtanko}, \citenamefont {Marvian}, \citenamefont {Lloyd}, \citenamefont {Ramanathan},\ and\ \citenamefont {Cappellaro}}]{Wei2019}%
  \BibitemOpen
  \bibfield  {author} {\bibinfo {author} {\bibfnamefont {K.~X.}\ \bibnamefont {Wei}}, \bibinfo {author} {\bibfnamefont {P.}~\bibnamefont {Peng}}, \bibinfo {author} {\bibfnamefont {O.}~\bibnamefont {Shtanko}}, \bibinfo {author} {\bibfnamefont {I.}~\bibnamefont {Marvian}}, \bibinfo {author} {\bibfnamefont {S.}~\bibnamefont {Lloyd}}, \bibinfo {author} {\bibfnamefont {C.}~\bibnamefont {Ramanathan}},\ and\ \bibinfo {author} {\bibfnamefont {P.}~\bibnamefont {Cappellaro}},\ }\bibfield  {title} {\bibinfo {title} {Emergent prethermalization signatures in out-of-time ordered correlations},\ }\href {https://doi.org/10.1103/PhysRevLett.123.090605} {\bibfield  {journal} {\bibinfo  {journal} {Phys. Rev. Lett.}\ }\textbf {\bibinfo {volume} {123}},\ \bibinfo {pages} {090605} (\bibinfo {year} {2019})}\BibitemShut {NoStop}%
\bibitem [{\citenamefont {Peng}\ \emph {et~al.}(2021)\citenamefont {Peng}, \citenamefont {Yin}, \citenamefont {Huang}, \citenamefont {Ramanathan},\ and\ \citenamefont {Cappellaro}}]{Peng2021}%
  \BibitemOpen
  \bibfield  {author} {\bibinfo {author} {\bibfnamefont {P.}~\bibnamefont {Peng}}, \bibinfo {author} {\bibfnamefont {C.}~\bibnamefont {Yin}}, \bibinfo {author} {\bibfnamefont {X.}~\bibnamefont {Huang}}, \bibinfo {author} {\bibfnamefont {C.}~\bibnamefont {Ramanathan}},\ and\ \bibinfo {author} {\bibfnamefont {P.}~\bibnamefont {Cappellaro}},\ }\bibfield  {title} {\bibinfo {title} {Floquet prethermalization in dipolar spin chains},\ }\href {https://doi.org/10.1038/s41567-020-01120-z} {\bibfield  {journal} {\bibinfo  {journal} {Nat. Phys.}\ }\textbf {\bibinfo {volume} {17}},\ \bibinfo {pages} {444} (\bibinfo {year} {2021})}\BibitemShut {NoStop}%
\bibitem [{\citenamefont {Rubio-Abadal}\ \emph {et~al.}(2020)\citenamefont {Rubio-Abadal}, \citenamefont {Ippoliti}, \citenamefont {Hollerith}, \citenamefont {Wei}, \citenamefont {Rui}, \citenamefont {Sondhi}, \citenamefont {Khemani}, \citenamefont {Gross},\ and\ \citenamefont {Bloch}}]{Rubio2020}%
  \BibitemOpen
  \bibfield  {author} {\bibinfo {author} {\bibfnamefont {A.}~\bibnamefont {Rubio-Abadal}}, \bibinfo {author} {\bibfnamefont {M.}~\bibnamefont {Ippoliti}}, \bibinfo {author} {\bibfnamefont {S.}~\bibnamefont {Hollerith}}, \bibinfo {author} {\bibfnamefont {D.}~\bibnamefont {Wei}}, \bibinfo {author} {\bibfnamefont {J.}~\bibnamefont {Rui}}, \bibinfo {author} {\bibfnamefont {S.~L.}\ \bibnamefont {Sondhi}}, \bibinfo {author} {\bibfnamefont {V.}~\bibnamefont {Khemani}}, \bibinfo {author} {\bibfnamefont {C.}~\bibnamefont {Gross}},\ and\ \bibinfo {author} {\bibfnamefont {I.}~\bibnamefont {Bloch}},\ }\bibfield  {title} {\bibinfo {title} {Floquet prethermalization in a bose-hubbard system},\ }\href {https://doi.org/10.1103/PhysRevX.10.021044} {\bibfield  {journal} {\bibinfo  {journal} {Phys. Rev. X}\ }\textbf {\bibinfo {volume} {10}},\ \bibinfo {pages} {021044} (\bibinfo {year} {2020})}\BibitemShut {NoStop}%
\bibitem [{\citenamefont {Bao}\ \emph {et~al.}(2026)\citenamefont {Bao}, \citenamefont {Zhu}, \citenamefont {Liu}, \citenamefont {Song}, \citenamefont {Jin}, \citenamefont {Zhu}, \citenamefont {Gao}, \citenamefont {Zhang}, \citenamefont {Wang}, \citenamefont {Zou}, \citenamefont {Tan}, \citenamefont {Zhang}, \citenamefont {Cui}, \citenamefont {Shen}, \citenamefont {Zhong}, \citenamefont {He}, \citenamefont {Wang}, \citenamefont {Yang}, \citenamefont {Wang}, \citenamefont {Shen}, \citenamefont {Liu}, \citenamefont {Han}, \citenamefont {Wu}, \citenamefont {Deng}, \citenamefont {Dong}, \citenamefont {Zhang}, \citenamefont {Li}, \citenamefont {Wang}, \citenamefont {Song}, \citenamefont {Cheng}, \citenamefont {Mondaini}, \citenamefont {Guo}, \citenamefont {Huang},\ and\ \citenamefont {Wang}}]{Bao2026}%
  \BibitemOpen
  \bibfield  {author} {\bibinfo {author} {\bibfnamefont {Z.}~\bibnamefont {Bao}}, \bibinfo {author} {\bibfnamefont {Z.}~\bibnamefont {Zhu}}, \bibinfo {author} {\bibfnamefont {Y.-R.}\ \bibnamefont {Liu}}, \bibinfo {author} {\bibfnamefont {Z.}~\bibnamefont {Song}}, \bibinfo {author} {\bibfnamefont {F.}~\bibnamefont {Jin}}, \bibinfo {author} {\bibfnamefont {X.}~\bibnamefont {Zhu}}, \bibinfo {author} {\bibfnamefont {Y.}~\bibnamefont {Gao}}, \bibinfo {author} {\bibfnamefont {C.}~\bibnamefont {Zhang}}, \bibinfo {author} {\bibfnamefont {N.}~\bibnamefont {Wang}}, \bibinfo {author} {\bibfnamefont {Y.}~\bibnamefont {Zou}}, \bibinfo {author} {\bibfnamefont {Z.}~\bibnamefont {Tan}}, \bibinfo {author} {\bibfnamefont {A.}~\bibnamefont {Zhang}}, \bibinfo {author} {\bibfnamefont {Z.}~\bibnamefont {Cui}}, \bibinfo {author} {\bibfnamefont {F.}~\bibnamefont {Shen}}, \bibinfo {author} {\bibfnamefont {J.}~\bibnamefont {Zhong}}, \bibinfo {author} {\bibfnamefont {Y.}~\bibnamefont {He}}, \bibinfo {author} {\bibfnamefont {H.}~\bibnamefont {Wang}}, \bibinfo {author} {\bibfnamefont {J.-N.}\ \bibnamefont {Yang}}, \bibinfo {author} {\bibfnamefont {Y.}~\bibnamefont {Wang}}, \bibinfo {author} {\bibfnamefont {J.}~\bibnamefont {Shen}}, \bibinfo {author} {\bibfnamefont {G.}~\bibnamefont {Liu}}, \bibinfo {author} {\bibfnamefont {Y.}~\bibnamefont {Han}}, \bibinfo {author} {\bibfnamefont {Y.}~\bibnamefont {Wu}}, \bibinfo {author} {\bibfnamefont {J.}~\bibnamefont {Deng}}, \bibinfo {author} {\bibfnamefont {H.}~\bibnamefont {Dong}}, \bibinfo {author} {\bibfnamefont {P.}~\bibnamefont {Zhang}}, \bibinfo {author} {\bibfnamefont {H.}~\bibnamefont {Li}}, \bibinfo {author} {\bibfnamefont {Z.}~\bibnamefont {Wang}}, \bibinfo {author} {\bibfnamefont {C.}~\bibnamefont {Song}}, \bibinfo {author} {\bibfnamefont {C.}~\bibnamefont {Cheng}}, \bibinfo {author} {\bibfnamefont {R.}~\bibnamefont {Mondaini}}, \bibinfo {author} {\bibfnamefont {Q.}~\bibnamefont {Guo}}, \bibinfo {author} {\bibfnamefont {B.}~\bibnamefont
  {Huang}},\ and\ \bibinfo {author} {\bibfnamefont {H.}~\bibnamefont {Wang}},\ }\bibfield  {title} {\bibinfo {title} {Fock space prethermalization and time-crystalline order on a quantum processor},\ }\href {https://doi.org/10.1103/bq3c-c3d8} {\bibfield  {journal} {\bibinfo  {journal} {Phys. Rev. Lett.}\ }\textbf {\bibinfo {volume} {137}},\ \bibinfo {pages} {050407} (\bibinfo {year} {2026})}\BibitemShut {NoStop}%
\bibitem [{\citenamefont {Liu}\ \emph {et~al.}(2026)\citenamefont {Liu}, \citenamefont {Liu}, \citenamefont {Liang}, \citenamefont {Deng}, \citenamefont {Chen}, \citenamefont {Shi}, \citenamefont {Li}, \citenamefont {Zhang}, \citenamefont {Chen}, \citenamefont {Fang}, \citenamefont {Feng}, \citenamefont {Gu}, \citenamefont {He}, \citenamefont {Huang}, \citenamefont {Li}, \citenamefont {Liu}, \citenamefont {Li}, \citenamefont {Mei}, \citenamefont {Peng}, \citenamefont {Song}, \citenamefont {Wang}, \citenamefont {Wang}, \citenamefont {Wang}, \citenamefont {Xiao}, \citenamefont {Xu}, \citenamefont {Xu}, \citenamefont {Yan}, \citenamefont {Yu}, \citenamefont {Yuan}, \citenamefont {Zhang}, \citenamefont {Zhao}, \citenamefont {Zhao}, \citenamefont {Zhou}, \citenamefont {Wang}, \citenamefont {Song}, \citenamefont {Tian}, \citenamefont {Mintert}, \citenamefont {Knolle}, \citenamefont {Moessner}, \citenamefont {Zhang}, \citenamefont {Zhang}, \citenamefont {Xiang}, \citenamefont {Zheng}, \citenamefont {Xu}, \citenamefont {Zhao},\ and\ \citenamefont {Fan}}]{Liu2026}%
  \BibitemOpen
  \bibfield  {author} {\bibinfo {author} {\bibfnamefont {Z.-H.}\ \bibnamefont {Liu}}, \bibinfo {author} {\bibfnamefont {Y.}~\bibnamefont {Liu}}, \bibinfo {author} {\bibfnamefont {G.-H.}\ \bibnamefont {Liang}}, \bibinfo {author} {\bibfnamefont {C.-L.}\ \bibnamefont {Deng}}, \bibinfo {author} {\bibfnamefont {K.}~\bibnamefont {Chen}}, \bibinfo {author} {\bibfnamefont {Y.-H.}\ \bibnamefont {Shi}}, \bibinfo {author} {\bibfnamefont {T.-M.}\ \bibnamefont {Li}}, \bibinfo {author} {\bibfnamefont {L.}~\bibnamefont {Zhang}}, \bibinfo {author} {\bibfnamefont {B.-J.}\ \bibnamefont {Chen}}, \bibinfo {author} {\bibfnamefont {C.-P.}\ \bibnamefont {Fang}}, \bibinfo {author} {\bibfnamefont {D.}~\bibnamefont {Feng}}, \bibinfo {author} {\bibfnamefont {X.-Y.}\ \bibnamefont {Gu}}, \bibinfo {author} {\bibfnamefont {Y.}~\bibnamefont {He}}, \bibinfo {author} {\bibfnamefont {K.}~\bibnamefont {Huang}}, \bibinfo {author} {\bibfnamefont {H.}~\bibnamefont {Li}}, \bibinfo {author} {\bibfnamefont {H.-T.}\ \bibnamefont {Liu}}, \bibinfo {author} {\bibfnamefont {L.}~\bibnamefont {Li}}, \bibinfo {author} {\bibfnamefont {Z.-Y.}\ \bibnamefont {Mei}}, \bibinfo {author} {\bibfnamefont {Z.-Y.}\ \bibnamefont {Peng}}, \bibinfo {author} {\bibfnamefont {J.-C.}\ \bibnamefont {Song}}, \bibinfo {author} {\bibfnamefont {M.-C.}\ \bibnamefont {Wang}}, \bibinfo {author} {\bibfnamefont {S.-L.}\ \bibnamefont {Wang}}, \bibinfo {author} {\bibfnamefont {Z.}~\bibnamefont {Wang}}, \bibinfo {author} {\bibfnamefont {Y.}~\bibnamefont {Xiao}}, \bibinfo {author} {\bibfnamefont {M.}~\bibnamefont {Xu}}, \bibinfo {author} {\bibfnamefont {Y.-S.}\ \bibnamefont {Xu}}, \bibinfo {author} {\bibfnamefont {Y.}~\bibnamefont {Yan}}, \bibinfo {author} {\bibfnamefont {Y.-H.}\ \bibnamefont {Yu}}, \bibinfo {author} {\bibfnamefont {W.-P.}\ \bibnamefont {Yuan}}, \bibinfo {author} {\bibfnamefont {J.-C.}\ \bibnamefont {Zhang}}, \bibinfo {author} {\bibfnamefont {J.-J.}\ \bibnamefont {Zhao}}, \bibinfo {author} {\bibfnamefont {K.}~\bibnamefont {Zhao}},
  \bibinfo {author} {\bibfnamefont {S.-Y.}\ \bibnamefont {Zhou}}, \bibinfo {author} {\bibfnamefont {Z.-A.}\ \bibnamefont {Wang}}, \bibinfo {author} {\bibfnamefont {X.}~\bibnamefont {Song}}, \bibinfo {author} {\bibfnamefont {Y.}~\bibnamefont {Tian}}, \bibinfo {author} {\bibfnamefont {F.}~\bibnamefont {Mintert}}, \bibinfo {author} {\bibfnamefont {J.}~\bibnamefont {Knolle}}, \bibinfo {author} {\bibfnamefont {R.}~\bibnamefont {Moessner}}, \bibinfo {author} {\bibfnamefont {Y.-R.}\ \bibnamefont {Zhang}}, \bibinfo {author} {\bibfnamefont {P.}~\bibnamefont {Zhang}}, \bibinfo {author} {\bibfnamefont {Z.}~\bibnamefont {Xiang}}, \bibinfo {author} {\bibfnamefont {D.}~\bibnamefont {Zheng}}, \bibinfo {author} {\bibfnamefont {K.}~\bibnamefont {Xu}}, \bibinfo {author} {\bibfnamefont {H.}~\bibnamefont {Zhao}},\ and\ \bibinfo {author} {\bibfnamefont {H.}~\bibnamefont {Fan}},\ }\bibfield  {title} {\bibinfo {title} {Prethermalization by random multipolar driving on a 78-qubit processor},\ }\href {https://doi.org/10.1038/s41586-025-09977-x} {\bibfield  {journal} {\bibinfo  {journal} {Nature}\ }\textbf {\bibinfo {volume} {650}},\ \bibinfo {pages} {79} (\bibinfo {year} {2026})}\BibitemShut {NoStop}%
\bibitem [{\citenamefont {Jurcevic}\ \emph {et~al.}(2014)\citenamefont {Jurcevic}, \citenamefont {Lanyon}, \citenamefont {Hauke}, \citenamefont {Hempel}, \citenamefont {Zoller}, \citenamefont {Blatt},\ and\ \citenamefont {Roos}}]{Jurcevic2014}%
  \BibitemOpen
  \bibfield  {author} {\bibinfo {author} {\bibfnamefont {P.}~\bibnamefont {Jurcevic}}, \bibinfo {author} {\bibfnamefont {B.~P.}\ \bibnamefont {Lanyon}}, \bibinfo {author} {\bibfnamefont {P.}~\bibnamefont {Hauke}}, \bibinfo {author} {\bibfnamefont {C.}~\bibnamefont {Hempel}}, \bibinfo {author} {\bibfnamefont {P.}~\bibnamefont {Zoller}}, \bibinfo {author} {\bibfnamefont {R.}~\bibnamefont {Blatt}},\ and\ \bibinfo {author} {\bibfnamefont {C.~F.}\ \bibnamefont {Roos}},\ }\bibfield  {title} {\bibinfo {title} {Quasiparticle engineering and entanglement propagation in a quantum many-body system},\ }\href@noop {} {\bibfield  {journal} {\bibinfo  {journal} {Nature}\ }\textbf {\bibinfo {volume} {511}},\ \bibinfo {pages} {202} (\bibinfo {year} {2014})}\BibitemShut {NoStop}%
\bibitem [{\citenamefont {Richerme}\ \emph {et~al.}(2014)\citenamefont {Richerme}, \citenamefont {Gong}, \citenamefont {Lee}, \citenamefont {Senko}, \citenamefont {Smith}, \citenamefont {Foss-Feig}, \citenamefont {Michalakis}, \citenamefont {Gorshkov},\ and\ \citenamefont {Monroe}}]{Richerme2014}%
  \BibitemOpen
  \bibfield  {author} {\bibinfo {author} {\bibfnamefont {P.}~\bibnamefont {Richerme}}, \bibinfo {author} {\bibfnamefont {Z.-X.}\ \bibnamefont {Gong}}, \bibinfo {author} {\bibfnamefont {A.}~\bibnamefont {Lee}}, \bibinfo {author} {\bibfnamefont {C.}~\bibnamefont {Senko}}, \bibinfo {author} {\bibfnamefont {J.}~\bibnamefont {Smith}}, \bibinfo {author} {\bibfnamefont {M.}~\bibnamefont {Foss-Feig}}, \bibinfo {author} {\bibfnamefont {S.}~\bibnamefont {Michalakis}}, \bibinfo {author} {\bibfnamefont {A.~V.}\ \bibnamefont {Gorshkov}},\ and\ \bibinfo {author} {\bibfnamefont {C.}~\bibnamefont {Monroe}},\ }\bibfield  {title} {\bibinfo {title} {Non-local propagation of correlations in quantum systems with long-range interactions},\ }\href {https://doi.org/10.1038/nature13450} {\bibfield  {journal} {\bibinfo  {journal} {Nature}\ }\textbf {\bibinfo {volume} {511}},\ \bibinfo {pages} {198} (\bibinfo {year} {2014})}\BibitemShut {NoStop}%
\bibitem [{\citenamefont {Li}\ \emph {et~al.}(2023)\citenamefont {Li}, \citenamefont {null}, \citenamefont {Colombo}, \citenamefont {Shu}, \citenamefont {Velez}, \citenamefont {Pilatowsky-Cameo}, \citenamefont {Schmied}, \citenamefont {Choi}, \citenamefont {Lukin}, \citenamefont {nafiel},\ and\ \citenamefont {Vuleti\'{c}}}]{Li2023}%
  \BibitemOpen
  \bibfield  {author} {\bibinfo {author} {\bibfnamefont {Z.}~\bibnamefont {Li}}, \bibinfo {author} {\bibnamefont {null}}, \bibinfo {author} {\bibfnamefont {S.}~\bibnamefont {Colombo}}, \bibinfo {author} {\bibfnamefont {C.}~\bibnamefont {Shu}}, \bibinfo {author} {\bibfnamefont {G.}~\bibnamefont {Velez}}, \bibinfo {author} {\bibfnamefont {S.}~\bibnamefont {Pilatowsky-Cameo}}, \bibinfo {author} {\bibfnamefont {R.}~\bibnamefont {Schmied}}, \bibinfo {author} {\bibfnamefont {S.}~\bibnamefont {Choi}}, \bibinfo {author} {\bibfnamefont {M.}~\bibnamefont {Lukin}}, \bibinfo {author} {\bibfnamefont {E.~P.-P.}\ \bibnamefont {nafiel}},\ and\ \bibinfo {author} {\bibfnamefont {V.}~\bibnamefont {Vuleti\'{c}}},\ }\bibfield  {title} {\bibinfo {title} {Improving metrology with quantum scrambling},\ }\href {https://doi.org/10.1126/science.adg9500} {\bibfield  {journal} {\bibinfo  {journal} {Science}\ }\textbf {\bibinfo {volume} {380}},\ \bibinfo {pages} {1381} (\bibinfo {year} {2023})}\BibitemShut {NoStop}%
\bibitem [{\citenamefont {Luo}\ \emph {et~al.}(2025)\citenamefont {Luo}, \citenamefont {Zhang}, \citenamefont {Chu}, \citenamefont {Maruko}, \citenamefont {Rey},\ and\ \citenamefont {Thompson}}]{Luo2025}%
  \BibitemOpen
  \bibfield  {author} {\bibinfo {author} {\bibfnamefont {C.}~\bibnamefont {Luo}}, \bibinfo {author} {\bibfnamefont {H.}~\bibnamefont {Zhang}}, \bibinfo {author} {\bibfnamefont {A.}~\bibnamefont {Chu}}, \bibinfo {author} {\bibfnamefont {C.}~\bibnamefont {Maruko}}, \bibinfo {author} {\bibfnamefont {A.~M.}\ \bibnamefont {Rey}},\ and\ \bibinfo {author} {\bibfnamefont {J.~K.}\ \bibnamefont {Thompson}},\ }\bibfield  {title} {\bibinfo {title} {Hamiltonian engineering of collective {XYZ} spin models in an optical cavity},\ }\href {https://doi.org/10.1038/s41567-025-02866-0} {\bibfield  {journal} {\bibinfo  {journal} {Nat. Phys.}\ }\textbf {\bibinfo {volume} {21}},\ \bibinfo {pages} {916} (\bibinfo {year} {2025})}\BibitemShut {NoStop}%
\bibitem [{\citenamefont {Russomanno}\ \emph {et~al.}(2021)\citenamefont {Russomanno}, \citenamefont {Fava},\ and\ \citenamefont {Heyl}}]{Russomanno2021}%
  \BibitemOpen
  \bibfield  {author} {\bibinfo {author} {\bibfnamefont {A.}~\bibnamefont {Russomanno}}, \bibinfo {author} {\bibfnamefont {M.}~\bibnamefont {Fava}},\ and\ \bibinfo {author} {\bibfnamefont {M.}~\bibnamefont {Heyl}},\ }\bibfield  {title} {\bibinfo {title} {Quantum chaos and ensemble inequivalence of quantum long-range ising chains},\ }\href {https://doi.org/10.1103/PhysRevB.104.094309} {\bibfield  {journal} {\bibinfo  {journal} {Phys. Rev. B}\ }\textbf {\bibinfo {volume} {104}},\ \bibinfo {pages} {094309} (\bibinfo {year} {2021})}\BibitemShut {NoStop}%
\bibitem [{\citenamefont {Pal}\ and\ \citenamefont {Santos}(2025)}]{Pal2025}%
  \BibitemOpen
  \bibfield  {author} {\bibinfo {author} {\bibfnamefont {S.~K.}\ \bibnamefont {Pal}}\ and\ \bibinfo {author} {\bibfnamefont {L.~F.}\ \bibnamefont {Santos}},\ }\href {https://arxiv.org/abs/2508.00077} {\bibinfo {title} {Fragmented eigenstate thermalization versus robust integrability in long-range models}} (\bibinfo {year} {2025}),\ \Eprint {https://arxiv.org/abs/2508.00077} {arXiv:2508.00077 [cond-mat.stat-mech]} \BibitemShut {NoStop}%
\bibitem [{\citenamefont {Sriram}\ \emph {et~al.}(2026)\citenamefont {Sriram}, \citenamefont {Pal},\ and\ \citenamefont {Santos}}]{Sriram2026}%
  \BibitemOpen
  \bibfield  {author} {\bibinfo {author} {\bibfnamefont {C.~L.}\ \bibnamefont {Sriram}}, \bibinfo {author} {\bibfnamefont {S.~K.}\ \bibnamefont {Pal}},\ and\ \bibinfo {author} {\bibfnamefont {L.~F.}\ \bibnamefont {Santos}},\ }\href {https://arxiv.org/abs/2607.15350} {\bibinfo {title} {Fragmented eth: Prethermalization, timescales, and ensemble inequivalence}} (\bibinfo {year} {2026}),\ \Eprint {https://arxiv.org/abs/2607.15350} {arXiv:2607.15350 [cond-mat.stat-mech]} \BibitemShut {NoStop}%
\bibitem [{not()}]{noteSUPPL}%
  \BibitemOpen
  \href@noop {} {}\bibinfo {note} {See Supplemental Material.}\BibitemShut {Stop}%
\end{thebibliography}%
%%%%%%%%%%%%%%%%%%%%%%%%%%%%%%%%%%%%%%%%%%%%%%%%%%%%%%%%%%%%%

%%%%%%%%%%%%%%%%%% SM %%%%%%%%%%%%%%%%%%%%%%%%%%%%%%

\onecolumngrid

\vspace*{0.5cm}

\begin{center}

{\large \bf Supplemental Material: 
\\Observable- and state-selective prethermalization and bounds on prethermal lifetimes}\\

\vspace{0.6cm}

C. L. Sriram$^1$, Soumya Kanti Pal$^2$, Lea F. Santos$^1$\\

$^1${\it Department of Physics, University of Connecticut, Storrs, Connecticut 06269, USA}

$^2${\it Department of Theoretical Physics, Tata Institute of Fundamental Research, Homi Bhabha Road, Mumbai 400005, India}

\end{center}

\vspace{0.6cm}

%%%\twocolumngrid

%\renewcommand{\theequation}{S\arabic{equation}}
%\renewcommand{\thefigure}{S\arabic{figure}}
\setcounter{secnumdepth}{2}
\setcounter{section}{0}
\renewcommand\thesection{\Roman{section}}
\renewcommand\thesubsection{\thesection.\Alph{subsection}}
\renewcommand{\thefigure}{S\arabic{figure}}
\renewcommand{\theequation}{S\arabic{equation}}

This Supplemental Material provides details of the derivation of the bound on prethermal lifetimes. It is organized as follows. Section~\ref{App:Proof_Obs} explains the absence of prethermal plateaus for observables with full permutation symmetry. Section~\ref{app:expansion} justifies why $\hat\rho_{\rm pre}(t)$ describes the dynamics up to the prethermal plateau. Section~\ref{APP:Trace-Norm} derives the trace-norm relation underlying the bound, and Sec.~\ref{APP:prethermal-timscale-proof} provides the detailed proof that the Loschmidt echo sets a lower bound on prethermal lifetimes. The proof is extended to initial mixed states and time-dependent Hamiltonians.
%%%%%%%%%%%%%%%%%%

\section{Permutation-invariant observables and the absence of prethermal plateaus}
\label{App:Proof_Obs}

To establish the absence of the prethermal plateau for a broad class of observables, we first characterize the structure of permutation-invariant operators in the eigenbasis of the unperturbed all-to-all Hamiltonian $\hat{H}^{(0)}$.

\textit{\textbf{Hilbert-space decomposition.}}---
Let
\begin{equation}
\mathcal H
=
\left(\frac{1}{2}\right)^{\otimes L}
\end{equation}
denote the Hilbert space of $L$ spin-$1/2$ particles. For even $L$, it decomposes into total-spin sectors as
\begin{align}
\mathcal H
=
\bigoplus_{s=0}^{L/2}
\left(
\mathcal V_s\otimes \mathcal M_s
\right),
\label{eq:HilbertDecomposition}
\end{align}
where $\mathcal V_s$ is the spin-$s$ irreducible representation of $SU(2)$, with dimension
\begin{align}
\dim(\mathcal V_s)=2s+1,
\end{align}
and $\mathcal M_s$ is the corresponding multiplicity space. Its dimension equals the degeneracy of each energy band,
\begin{align}
\dim(\mathcal M_s)
=g(s,L) =  \frac{2s + 1}{L + 1} \binom{L + 1}{\frac{L}{2} - s},
\end{align}
which follows from the combinatorics of angular momentum addition. The energy bands are highly degenerate, except for the fully symmetric sector,
$s=L/2$. In this case,
$g(L/2,L)=1$, so each of the $(L+1)$
energy bands contains a single eigenstate. 

The fully connected Hamiltonian is invariant under arbitrary permutations of the spins. As a consequence, it acts nontrivially only on the spin-$s$ representation and trivially on the multiplicity space,
\begin{align}
\hat H^{(0)}
=
\bigoplus_{s=0}^{L/2}
\left(
\hat H_s^{(0)}
\otimes
\hat{\mathbb I}_{\mathcal M_s}
\right).
\label{eq:H0Block}
\end{align}

Therefore, the eigenstates factorize as
\begin{align}
|E_{n,s}^{(0)},\lambda\rangle
=
|\phi_n^{(s)}\rangle
\otimes
|\lambda\rangle ,
\label{eq:FactorizedEigenstates}
\end{align}
where
\begin{align}
\hat H_s^{(0)}
|\phi_n^{(s)}\rangle
=
E_{n,s}^{(0)}
|\phi_n^{(s)}\rangle ,
\end{align}
and
\begin{align}
|\phi_n^{(s)}\rangle
=
\sum_{m=-s}^{s}
c_{m,n}^{(s)}
|s,m\rangle .
\end{align}
In the equations above, $n$ labels the eigenstate inside the spin-$s$ representation $\mathcal V_s$. This determines the energy. On the other hand, $\lambda$ labels the different copies (multiplicity) of the same spin-$s$ representation. These copies are completely equivalent from the point of view of any permutation-invariant Hamiltonian. Therefore, the energy depends only on $n$ and $s$.

\textit{\textbf{Lemma.}}---
Let $\hat O:\mathcal H\rightarrow\mathcal H$ be invariant under arbitrary permutations of the spin sites,
\begin{align}
[\hat O,\hat U_\pi]=0,
\qquad
\forall \pi\in S_L,
\end{align}
where $S_L$ denotes the symmetric group of $L$ objects and $\hat U_\pi$ is the unitary operator implementing the permutation $\pi$. 

As for the fully connected Hamiltonian, permutation invariance implies that
$\hat O$ cannot distinguish between the different copies of a given
spin-$s$ irreducible representation labeled by the multiplicity index
$\lambda$. Consequently, it acts nontrivially only on the spin
representation $\mathcal V_s$ and as the identity on the multiplicity
space $\mathcal M_s$. Therefore,
\begin{align}
\hat O
=
\bigoplus_s
\left(
\hat o_s
\otimes
\hat{\mathbb I}_{\mathcal M_s}
\right),
\label{eq:ObservableBlock}
\end{align}
and its matrix elements in the eigenbasis of $\hat H^{(0)}$ satisfy
\begin{align}
\langle E_{n,s}^{(0)},\lambda|
\hat O
|E_{n',s'}^{(0)},\lambda'\rangle =
\langle \phi_n^{(s)}
|
\hat o_s
|
\phi_{n'}^{(s)}
\rangle
\delta_{s s'}
\delta_{\lambda \lambda'} .
\label{eq:SelectionRule}
\end{align}
Equation~(\ref{eq:SelectionRule}) implies that matrix elements between states with different multiplicity indices vanish.

\textit{\textbf{Proof.}}---
Under the action of the permutation group $S_L$, the multiplicity spaces $\mathcal M_s$ carry inequivalent irreducible representations. By Schur's lemma, any operator commuting with all permutations must act as the identity on each multiplicity space. Therefore,
\begin{align}
\hat O
=
\bigoplus_s
\left(
\hat o_s
\otimes
\hat{\mathbb I}_{\mathcal M_s}
\right),
\end{align}
which establishes Eq.~\eqref{eq:ObservableBlock}.

Using the factorized eigenstates in Eq.~\eqref{eq:FactorizedEigenstates},
\begin{align}
&\langle E_{n,s}^{(0)},\lambda|
\hat O
|E_{n',s'}^{(0)},\lambda'\rangle =
(\langle\phi_n^{(s)}|\otimes\langle\lambda|)
(\hat o_s\otimes\hat{\mathbb I}_{\mathcal M_s})
(|\phi_{n'}^{(s')}\rangle\otimes|\lambda'\rangle)
=
\langle\phi_n^{(s)}|
\hat o_s
|\phi_{n'}^{(s)}\rangle
\delta_{\lambda\lambda'} \delta_{ss'},
\end{align}
which proves Eq.~\eqref{eq:SelectionRule}.

The first term in the following equation given in the main text,
\begin{align}
    \langle \hat{O} \rangle (t) &= \!\! \!\! \sum_{\substack{m \neq n \\ E_m^{(0)} \neq E_n^{(0)}}} \!\!\!\! \! \!{c_m^*}^{(0)} c_n^{(0)} \langle m^{(0)}|\hat{O}|n^{(0)} \rangle e^{i t[(E_m^{(0)} - E_n^{(0)}) 
    + \mathcal{O}(\epsilon)]} \nonumber \\
    & + \!\!\!\! \sum_{\substack{m \neq n \\ E_m^{(0)} = E_n^{(0)}}} \!\! \!\! \! \! {c_m^*}^{(0)} c_n^{(0)} \langle m^{(0)}|\hat{O}|n^{(0)} \rangle e^{i\epsilon t(E_m^{(1)} - E_n^{(1)})} \nonumber \\
    & + \sum_n |c_n^{(0)}|^2\langle n^{(0)}|\hat{O}|n^{(0)} \rangle  ,
\end{align}
originates from coherences between different states that remain degenerate under the LMG Hamiltonian. These states differ only by their multiplicity label $\lambda$. Equation~(\ref{eq:SelectionRule}) shows that a permutation-invariant observable has vanishing matrix elements between different multiplicity sectors. Consequently, the first term in the equation above is identically zero, so permutation-invariant observables do not exhibit distinguishable prethermal plateaus.

%%%%%%%%%%%%%%%%%
\section{Short-time expansion of the evolution operator}
\label{app:expansion}

In this section, we justify why
\begin{align}
\hat\rho_{\rm pre}(t) = e^{-i\hat H_0t} \hat\rho_0 e^{i\hat H_0t},
\label{Eq:rhoPRE}
\end{align}
describes the dynamics up to the prethermal regime for any Hamiltonian of the form
\begin{align}
\hat H=\hat H_0+\epsilon \hat V,
\end{align}
where $\epsilon\ll1$ controls the strength of the perturbation.

To this end, consider two generally noncommuting operators $A$ and $B$ and define
\begin{align}
F(s)=e^{-sA}e^{s(A+\epsilon B)},
\qquad
F(0)=\mathbb I .
\end{align}
Differentiating with respect to $s$ gives
\begin{align}
\frac{dF}{ds}
&=
-Ae^{-sA}e^{s(A+\epsilon B)}
+e^{-sA}(A+\epsilon B)e^{s(A+\epsilon B)}
=
\epsilon\,e^{-sA}Be^{sA}F(s).
\end{align}
Since the generators
$e^{-sA}Be^{sA}$
at different values of $s$ generally do not commute, the solution is the ordered exponential
\begin{align}
F(s)
=
\mathcal T
\exp\!\left[
\epsilon
\int_0^s
ds'\,
e^{-s'A}
Be^{s'A}
\right],
\end{align}
where $\mathcal T$ denotes ordering with respect to the parameter $s'$.

Setting $s=1$ yields the exact operator identity
\begin{align}
e^{A+\epsilon B}
=
e^A
\,
\mathcal T
\exp\!\left[
\epsilon
\int_0^1
ds\,
e^{-sA}
Be^{sA}
\right].
\label{eq:ordered_identity}
\end{align}

We now apply Eq.~\eqref{eq:ordered_identity} to the time-evolution operator by choosing
\begin{align}
A=-it\hat H_0,
\qquad
B=-it\hat V,
\end{align}
which gives
\begin{align}
e^{-it(\hat H_0+\epsilon\hat V)}
=
e^{-it\hat H_0}
\mathcal T
\exp\!\left[
-i\epsilon t
\int_0^1
ds\,
e^{is t\hat H_0}
\hat V
e^{-is t\hat H_0}
\right].
\label{eq:Uexact}
\end{align}

For times satisfying
\begin{align}
\epsilon t\ll1,
\end{align}
the ordered exponential can be expanded to first order,
\begin{align}
& e^{-it(\hat H_0+\epsilon\hat V)} =
e^{-it\hat H_0}
\left[
1
-
i\epsilon t
\int_0^1
ds\,
e^{is t\hat H_0}
\hat V
e^{-is t\hat H_0}
+
\mathcal O(\epsilon^2t^2)
\right].
\label{eq:Upert}
\end{align}
Equation~\eqref{eq:Upert} shows that, up to corrections of order $\epsilon t$, the exact evolution operator coincides with that generated by the parent Hamiltonian $\hat H_0$. Consequently, for times much shorter than the inverse perturbation strength,
\begin{align}
t\ll\epsilon^{-1},
\end{align}
the density matrix evolves as
\begin{align}
\hat\rho(t)
=
e^{-it\hat H}\hat\rho_0e^{it\hat H}
=
e^{-it\hat H_0}
\hat\rho_0
e^{it\hat H_0}
+
\mathcal O(\epsilon t),
\end{align}
which justifies the definition of the prethermal evolution used in the main text,
\begin{align}
\hat\rho_{\rm pre}(t)
=
e^{-it\hat H_0}
\hat\rho_0
e^{it\hat H_0}.
\end{align}

\section{Trace-norm inequality} \label{APP:Trace-Norm}
\textit{\textbf{Lemma}}:---
Let $A$ be a bounded operator and $X$ be a trace-class operator on a finite-dimensional Hilbert space. Then
\begin{equation}
|\mathrm{tr}(A X)| \le \|A\|_{\infty} ~ \|X\|_1,
\end{equation}
where $\|A\|_{\infty}$ is the operator norm and $\|X\|_1 = \text{Tr}[\sqrt{X^\dagger X}]$ is the trace norm. The operator norm is the largest singular value and for a positive semidefinite Hermitian operator, it coincides with the largest eigenvalue.

\textit{\textbf{Proof}}:---
Let 
\begin{equation}
X = U \Sigma V^\dagger,
\end{equation}
be the singular value decomposition of $X$, 
where $U$ and $V$ are unitary matrices, and $\Sigma = \mathrm{diag}(s_1, s_2, \dots, s_n)$ with $s_i \ge 0$ the singular values of $X$. Then
\begin{align}
|\mathrm{tr}(A X)| &= |\mathrm{tr}(A U \Sigma V^\dagger)| = |\mathrm{tr}(V^\dagger A U \, \Sigma)| = \left| \sum_i s_i (V^\dagger A U)_{ii} \right|.
\end{align}

Applying the triangle inequality and using the definition of the operator norm $\|W\|_{\infty} := \sup_{\|v\|=1} \|W v\|$, we get
\begin{equation}
\left| \sum_i s_i (V^\dagger A U)_{ii} \right| \le \sum_i s_i |(V^\dagger A U)_{ii}| \le \sum_i s_i \, \|V^\dagger A U\|_{\infty}.
\end{equation}

Since the operator norm is unitarily invariant,
\begin{equation}
\|V^\dagger A U\|_{\infty} = \|A\|_{\infty},
\end{equation}
we obtain
\begin{equation}
|\mathrm{tr}(A X)| \le \sum_i s_i \, \|A\|_{\infty} = \|A\|_{\infty} \sum_i s_i = \|A\|_{\infty} ~ \|X\|_1.
\end{equation}

This completes the proof.

\section{Proof of the Loschmidt-echo lower bound on prethermal lifetimes} \label{APP:prethermal-timscale-proof}

For an initial pure state $| \Psi_0 \rangle$ with the corresponding density matrix $\hat\rho_0 = | \Psi_0 \rangle \langle \Psi_0|  $, we consider its evolution under two unitary operators $\hat U_0(t) = e^{-i\hat H_0 t  }$ and $\hat U(t) =e^{-i (\hat H_0 + \epsilon \hat V)  t  }$, generated by the Hamiltonian $\hat{H}_0$, and its weakly perturbed, non-integrable counterpart  $\hat H = \hat{H}_0 + \epsilon \hat V$, respectively. Next, for a bounded local observable $\hat O$, we define the instantaneous absolute deviation between its expectation values under the perturbed and prethermal (integrable) dynamics as 
\begin{align}
    & \chi_O(t) \equiv | \langle O \rangle(t) - \langle O \rangle_{\text{pre}}(t)|= \biggr|\text{Tr}[ \hat O (\hat{\rho}(t) - \hat{\rho}_{\text{pre}}(t) ) ]\biggr|, \label{eq:observable-deviation}
\end{align}
respectively, where
\begin{align}
    \hat\rho(t) = \hat U (t) \hat \rho_0 \hat U^\dagger (t) \qquad \text{and} \qquad \hat\rho_{\text{pre}}(t) = \hat U_0 (t) \hat \rho_0 \hat U^\dagger_0 (t)
\end{align}
denotes the density matrices evolved under the perturbed and the integrable Hamiltonians, respectively.

Moving forward, we apply the trace-norm inequality (see  Sec.~\ref{APP:Trace-Norm}) in Eq.~\eqref{eq:observable-deviation}, and subsequently obtain  
\begin{align} \label{Eq:App:observable-trace-norm}
    & \chi_O\le \| \hat{O}\|_{\infty} ~ \|(\hat\rho(t) - \hat\rho_{\text{pre}}(t))\|_1,
\end{align}
with $ \| \hat{O}\|_\infty = \max{\biggr(Spec(\hat O)\biggr)} $(maximum eigenvalue), and $\| A \|_1= \text{Tr}\biggr(\sqrt{A^\dagger A} \biggr)= \sum_{i}|\lambda_i|$ if $A=A^\dagger $, where $\lambda_i $'s are the eigenvalues of $\hat A$. 

Next, we establish that $\|(\hat\rho(t) - \hat\rho_{\text{pre}}(t))\|_1$ is related to the prethermal deviation of the Loschmidt echo. As stated in the main text, the Loschmidt echo is defined as  
\begin{align}
    \mathcal{L}(t) \equiv | \langle \psi_0 | \hat U_0^\dagger(t) \hat U(t) | \psi_0 \rangle|^2   =| \langle \psi_0 | e^{i \hat H_0 t} e^{-i \hat H t} | \psi_0 \rangle |^2.
\end{align}
We then define the corresponding prethermal deviation of the Loschmidt echo as
\begin{align} \label{eq:APP:loschmidt-echo-deviation}
    \chi_{\mathcal L}(t) &=  | \mathcal{L}^{\text{pre}}(t) - \mathcal{L}(t)| = 1-\mathcal L(t),
\end{align}
where the last equality follows from the fact that $\mathcal{L}^{\text{pre}}(t) = 1$. 

To proceed, we first observe that the density matrix difference $ \Delta\rho(t) \equiv \hat\rho(t) - \hat\rho_{\text{pre}}(t)$ can be written as the following 
\begin{align}
    \Delta\hat{\rho}(t)  = |\Psi (t) \rangle \langle \Psi (t) | - |\Psi (t) \rangle_{\text{pre}} (|\Psi (t) \rangle_{\text{pre}})^\dagger,
\end{align}
where $|\Psi (t) \rangle= \hat U(t) |\Psi_0\rangle$ and $|\Psi (t) \rangle_{\text{pre}}= \hat U_0(t) |\Psi_0\rangle \equiv |\phi(t) \rangle $. Since each projector is of rank one, the rank inequality $\text{Rank} (A + B) \le \mathrm{Rank}(A) + \mathrm{Rank}(B)$ immediately yields that  $\Delta \hat{\rho} (t)$ is at most of rank-$2$ for all times. Equivalently, $\Delta \hat{\rho} (t)$ has support only on the two-dimensional subspace spanned by $\{ | \Psi(t)\rangle , |\phi(t)\rangle \}$. Consequently, in a basis adapted to this subspace,
$\Delta \hat{\rho}(t)$
assumes the following block-diagonal structure: 
\begin{align}
    \Delta\rho(t)  =  \begin{pmatrix}
    (\Delta \rho(t))_{2\times2} & 0 \\
    0 & 0_{(\mathcal{D}-2) \times (\mathcal{D}-2) }.
\end{pmatrix},
\end{align} 
where $\mathcal{D} = 2^{L}$ is the dimension of the total Hilbert space.

Next, we explicitly construct the effective $2\times 2$ representation of $\Delta\hat\rho(t)$ by introducing an orthonormal basis $\{|e_1\rangle,|e_2\rangle\}$ for the subspace spanned by $\{|\Psi(t)\rangle,|\phi(t)\rangle\}$ via the Gram--Schmidt procedure:
\begin{align}
    & |e_1\rangle = | \Psi(t) \rangle,\\
    & | e_2\rangle = \frac{|\phi(t) \rangle - c| \Psi(t) \rangle  }{\sqrt{1 - |c |^2}},
\end{align}
with $c= \langle \psi(t) |\phi(t)\rangle $.
In this basis, the effective matrix becomes 
\begin{align}
    & \left(\Delta \hat{\rho} (t)\right)_{2\times 2}  = \begin{pmatrix}
       1- |c |^2  & -c\sqrt{1 - |c |^2}\\
        -c^*\sqrt{1 - |c |^2} & -(1- |c |^2)
    \end{pmatrix},
\end{align}
with the eigenvalues $\lambda_{\pm} = \pm \sqrt{1- |c|^2}$. Therefore, the trace norm of the density matrix difference is given exactly by
\begin{align}
    \|(\hat\rho(t) - \hat\rho_{\text{pre}}(t))\|_1 & = 2 \sqrt{1- |c|^2} = 2\sqrt{1- \mathcal{L}(t) },
\end{align}
where we have used $|c|^2 = \mathcal{L}(t)$. Plugging this into Eq.~\eqref{Eq:App:observable-trace-norm} yields 
\begin{align} \label{eq:LE-inequality}
    \chi_O\le 2 \| \hat{O}\|_\infty\sqrt{1- \mathcal{L}(t)} .
\end{align}

We now define the departure times from the prethermal plateau for the observable and the Loschmidt echo as
\begin{align}
    & t_O \equiv \inf \biggr[  t\ge 0 : \chi_O(t) \ge \delta_O  \biggr] \\
    & t_\mathcal{L} \equiv \inf \biggr[  t\ge 0 : \chi_{\mathcal{L}}(t)  \ge \delta_L \equiv \frac{\delta_O^2}{4 \| \hat{O}\|_\infty^2} \biggr].
\end{align}
For every time $t<t_{\mathcal L}$,
\begin{equation}
  \chi_{\mathcal L}(t) < \delta_{\mathcal L} .  
\end{equation}
Therefore
\begin{align}
\chi_O(t) \le 2 \| \hat{O}\|_\infty \sqrt{\chi_{\mathcal L}(t)}  < 2
\|\hat O\|_\infty
\sqrt{\delta_{\mathcal L}} =
\delta_O.
\end{align}
Hence, $t<t_O$ for every $t<t_{\mathcal L}$, which implies 
\begin{equation}
t_{\mathcal{L}} \le t_{O}, 
\end{equation}
which means that the Loschmidt echo departs from its prethermal plateau no later than the observable.

\subsection{Extension to mixed states}
\label{App:Mixed}
An important extension of the above proof is to consider the initial density matrix to be a mixed state $\hat\rho_0$ such that $\hat\rho_0^2 \neq \hat \rho_0 $. The discussion till Eq.~\eqref{eq:APP:loschmidt-echo-deviation} follows through, however $\Delta\rho(t) = \hat\rho(t) - \hat\rho_{\text{pre}}(t)$ is not a rank-$2$ matrix, and thereby, the bound in Eq.~\eqref{eq:LE-inequality} needs to be modified.

To this end, we use the following upper Fuchs-van de Graff inequality 
\begin{align}
    \| \hat\rho(t) - \hat\rho_{\text{pre}}(t) \|_1 \le 2 \sqrt{1 - \mathcal{F}(t)},
\end{align}
where $\mathcal{F}(t)$ is the Uhlmann fidelity, defined as 
\begin{align}
    \mathcal{F}(t) = \left( \text{Tr}\left[ \sqrt{\sqrt{\hat\rho(t) }\hat\rho_{\text{pre}(t) } \sqrt{\hat\rho(t) }} \right] \right)^2.
\end{align}
This immediately modifies Eq.~\eqref{eq:LE-inequality} to 
\begin{align}
    \chi_O\le 2 \sqrt{1- \mathcal{F}(t)} \| \hat{O}\|_\infty.
\end{align}
Defining
\begin{align}
\chi_{\mathcal F}(t)=1-\mathcal F(t)
\end{align}
and the corresponding departure threshold
\begin{align}
\delta_{\mathcal F}
=
\frac{\delta_O^2}{4\|\hat O\|_\infty^2},
\end{align}
the same argument as for pure states yields
\begin{align}
t_{\mathcal F}\leq t_O.
\end{align}
Thus, for mixed initial states, the fidelity between the exact and
prethermal dynamics provides the corresponding lower bound on the
prethermal lifetime of any bounded observable. For a pure initial state,
$\mathcal F(t)=\mathcal L(t)$, and the result reduces to the
Loschmidt-echo bound derived above.

%%%%%%%%%%%%%%%%%%%%%%%%
\subsection{Extension to time-dependent Hamiltonians}
\label{App:TimeDependent}

Neither the pure-state nor the mixed-state bounds derived above requires
the exact or prethermal evolution to be generated by a time-independent
Hamiltonian. Both results require only that the exact and prethermal
dynamics be unitary and therefore extend directly to time-dependent
Hamiltonians.

Consider a general time-dependent Hamiltonian $\hat H(t)$, with exact
evolution operator
\begin{equation}
\hat U(t,0)
=
\mathcal{T}
\exp\left[
-i\int_0^t ds,\hat H(s)
\right],
\end{equation}
where $\mathcal{T}$ denotes time ordering. Suppose that, within the
prethermal regime, the dynamics is described by a unitary reference
evolution $\hat U_{\rm pre}(t,0)$. For a pure initial state, the
generalized Loschmidt echo is
\begin{equation}
\mathcal{L}(t)
=
\left|
\langle\Psi_0|
\hat U_{\rm pre}^\dagger(t,0)
\hat U(t,0)
|\Psi_0\rangle
\right|^2,
\end{equation}
and the preceding proof immediately gives
\begin{equation}
t_{\mathcal L}\leq t_O.
\end{equation}
For a mixed initial state, the corresponding result follows from the
Uhlmann fidelity $\mathcal F(t)$ between the exact and prethermal
density matrices,
\begin{equation}
t_{\mathcal F}\leq t_O.
\end{equation}
Thus, the bounds derived above apply to arbitrary time-dependent
Hamiltonians whenever the exact and prethermal dynamics are unitary.

Periodically driven systems provide an important special case. For
$\hat H(t+T)=\hat H(t)$, a high-frequency Floquet prethermal evolution
may be written in terms of a prethermal effective Hamiltonian and a
periodic micromotion operator $\hat P(t)$ as
\begin{equation}
\hat U_{\rm pre}(t,0)
=
\hat P(t)e^{-i\hat H_{\rm eff}t}\hat P^\dagger(0),
\qquad
\hat P(t+T)=\hat P(t).
\end{equation}
The bounds therefore apply both at stroboscopic times and, when the
micromotion is retained in $\hat U_{\rm pre}(t,0)$, at arbitrary times.

\end{document}